\documentclass[dvipdfm]{aa}
\usepackage{graphicx,amssymb}
\usepackage[normalem]{ulem}
\usepackage{natbib}
\usepackage{lscape}

\bibpunct{(}{)}{;}{a}{}{,}

\begin{document}

%
%       NEW COMMANDS
%
\newcommand{\text}[1]{#1}
\newcommand{\wav}[1]{$\lambda#1\,\mathrm{cm}$}  %one wavelength
\newcommand{\wwav}[2]{$\lambda\lambda#1,#2\,\mathrm{cm}$}  %2 wavelengths
\newcommand{\wwwav}[3]{$\lambda\lambda#1,#2,#3\,\mathrm{cm}$}  %3 wavelengths
\newcommand{\Bt}{\,B_\mathrm{tot}} %total field
\newcommand{\Btpa}{\,B_\mathrm{tot\parallel}}
\newcommand{\Btpe}{\,B_\mathrm{tot\perp}}
\newcommand{\Br}{\,B} %regular field
\newcommand{\BrO}{\,B_0}
\newcommand{\Brpa}{\,B_{\parallel}}
\newcommand{\Brpe}{\,B_{\perp}}
\newcommand{\Bur}{\,B_{r}} %cylindrical components
\newcommand{\But}{\,B_{\theta}}
\newcommand{\Buz}{\,B_{z}}
\newcommand{\bt}{\,b} %turbulent field
\newcommand{\btO}{\,b_0}
\newcommand{\btpa}{\,b_{\parallel}}
\newcommand{\btpe}{\,b_{\perp}}
%various differentials
\newcommand{\pdif}{\partial}
\newcommand{\pdt}[1][long]{\ifthenelse{\equal{#1}{long}}
  {\frac{\partial}{\partial t}}{\frac{\partial #1}{\partial t}}}
\newcommand{\pdr}[1][long]{\ifthenelse{\equal{#1}{long}}
  {\frac{\partial}{\partial r}}{\frac{\partial #1}{\partial r}}}
\newcommand{\pddr}[1][long]{\ifthenelse{\equal{#1}{long}}
  {\frac{\partial^2}{\partial r^2}}{\frac{\partial^2 #1}{\partial r^2}}}
\newcommand{\pdp}[1][long]{\ifthenelse{\equal{#1}{long}}
  {\frac{\partial}{\partial \phi}}{\frac{\partial #1}{\partial \phi}}}
\newcommand{\pddp}[1][long]{\ifthenelse{\equal{#1}{long}}
  {\frac{\partial^2}{\partial \phi^2}}{\frac{\partial^2 #1}{\partial \phi^2}}}
\newcommand{\pdz}[1][long]{\ifthenelse{\equal{#1}{long}}
  {\frac{\partial}{\partial z}}{\frac{\partial #1}{\partial z}}}
\newcommand{\pddz}[1][long]{\ifthenelse{\equal{#1}{long}}
  {\frac{\partial^2}{\partial z^2}}{\frac{\partial^2 #1}{\partial z^2}}}
\newcommand{\dif}{\,\mathrm{d}}
\newcommand{\curl}{\nabla\times}
\newcommand{\Lap}{\nabla^2}
\newcommand{\HI}{\mathrm{H\,\scriptstyle I}}
\newcommand{\HII}{\mathrm{H\,\scriptstyle II}}
%
%       UNITS
%
\newcommand{\cm}{\,\mathrm{cm}}
\newcommand{\mm}{\,\mathrm{mm}}
\newcommand{\cmcube}{\,\mathrm{cm^{-3}}}
\newcommand{\dyn}{\,\mathrm{dyn}}
\newcommand{\erg}{\,\mathrm{erg}}
\newcommand{\FRM}{\,\mathrm{rad\,m^{-2}}} % rotation measure
\newcommand{\g}{\,\mathrm{g}}
\newcommand{\Hz}{\,\mathrm{Hz}}
\newcommand{\GHz}{\,\mathrm{GHz}}
\newcommand{\Jy}{\,\mathrm{Jy}} % Jansky
\newcommand{\Jyb}{\,\mathrm{Jy/beam}} % Jansky per beam
\newcommand{\cms}{\,\mathrm{cm\,s^{-2}}}
\newcommand{\kms}{\,\mathrm{km\,s^{-1}}}
\newcommand{\mJy}{\,\mathrm{mJy}} % milli-Jansky
\newcommand{\mJyb}{\,\mathrm{mJy/beam}} % milli -Jansky per beam
\newcommand{\K}{\,\mathrm{K}}
\newcommand{\kpc}{\,\mathrm{kpc}}
\newcommand{\Mpc}{\,\mathrm{Mpc}}
\newcommand{\mG}{\,\mathrm{mG}} % milli-Gauss
\newcommand{\MHz}{\, \mathrm{MHz}}
\newcommand{\Msol}{\,\mathrm{M_\sun}}
\newcommand{\n}{\,n_\mathrm{e}}
\newcommand{\pc}{\,\mathrm{pc}}
\newcommand{\RM}{\,\mathrm{RM}}
\newcommand{\RMi}{\,\mathrm{RM_i}} % intrinsic RM
\newcommand{\RMfg}{\,\mathrm{RM_{fg}}} % foreground RM
\newcommand{\s}{\,\mathrm{s}}
\newcommand{\micron}{\,\mu\mathrm{m}} % micron
\newcommand{\uG}{\,\mu\mathrm{G}} % micro-Gauss
\newcommand{\uJy}{\,\mu\mathrm{Jy}} % micro-Jansky
\newcommand{\uJyb}{\,\mu\mathrm{Jy/beam}} % micro-Jansky per beam
\newcommand{\yr}{\,\mathrm{yr}}

\title{Magnetic field structures of galaxies derived from analysis of
Faraday rotation measures, and perspectives for the SKA}

%\author{R.~Stepanov \inst{1}
%  \and T.G.~Arshakian \inst{2}
%  \and R.~Beck \inst{2}
%  \and P.~Frick \inst{1}
%  \and M.~Krause \inst{2}

\author{Rodion~Stepanov \inst{1}
  \and Tigran G.~Arshakian \inst{2}\fnmsep\thanks{On leave from
           Byurakan Astrophysical Observatory, Byurakan 378433,
           Armenia and Isaac Newton Institute of Chile, Armenian
           Branch}
  \and Rainer~Beck \inst{2}
  \and Peter~Frick \inst{1}
  \and Marita~Krause \inst{2}
}

\institute{Institute of Continuous Media Mechanics, Korolyov str.~1,
  614061 Perm, Russia
  \and Max-Planck-Institut f\"ur Radioastronomie, Auf dem H\"ugel 69,
  53121 Bonn, Germany}

\offprints{R.Stepanov, \email{rodion@icmm.ru}; R.Beck,
\email{rbeck@mpifr-bonn.mpg.de}}

\date{Received / Accepted }

\abstract
% Context
{The forthcoming new-generation radio telescope SKA (Square
Kilometre Array) and its precursors will provide a rapidly growing
number of polarized radio sources.}
% Aims
{Our analysis looks at what can be learned from these sources
concerning the structure and evolution of magnetic fields of
external galaxies.}
% Methods
{ \emph{Recognition} of magnetic structures is possible from Faraday
rotation measures (${\rm RM}$) towards background sources behind
galaxies or a continuous $\RM$ map obtained from the diffuse
polarized emission from the galaxy itself. We constructed models for
the ionized gas and magnetic field patterns of different azimuthal
symmetries (axisymmetric, bisymmetric and quadrisymmetric spirals,
and superpositions) plus a halo magnetic field. $\RM$ fluctuations
with a Kolmogorov spectrum due to turbulent fields and/or
fluctuations in ionized gas density are superimposed. Assuming
extrapolated number density counts of polarized sources, we
generated a sample of $\RM$ values within the solid angle of the
galaxy. Applying various templates, we derived the minimum number of
background sources and the minimum quality of the observations. For
a large number of sources, \emph{reconstruction} of the field
structure without precognition becomes possible.}
% Results
{Any large-scale regular component of the magnetic field can be
clearly recognized from $\RM$ data with the help of the $\chi^2$
criterium. Under favorite conditions, a few dozen polarized sources
are enough for a reliable result. A halo field with a vertical
component does not affect the results of recognition. The required
source number increases for small inclinations of the galaxy's disk
and for larger $\RM$ turbulence. A flat number density distribution
of the sources can be overcome by more sensitive observations.
Application of the recognition method to the available $\RM$ data in
the region around M~31 indicates that there are significant $\RM$
contributions intrinsic to the background sources or due to the
foreground of the Milky Way. A reliable reconstruction of the field
structure needs at least 20 $\RM$ values on a cut along the
projected minor axis.}
% Conclusions
{Recognition or reconstruction of regular field structures from the
$\RM$ data of polarized background sources is a powerful tool for
future radio telescopes. Measuring $\RM$ at frequencies around 1~GHz
with the SKA, simple field structures can be recognized in galaxies
up to about 100\,Mpc distance and will allow to test dynamo against
primordial or other models of field origin. The low-frequency SKA
array and low-frequency precursor telescopes like LOFAR may also
have good $\RM$ sensitivity if background sources are still
significantly polarized at low frequencies.}


\titlerunning{Reconstruction of magnetic field structures}
\authorrunning{Stepanov et al.}

\maketitle

% --------------------------------------------------------------------

\section{Introduction }
\label{sec:intro}

Magnetic fields play an important role for the structure and
evolution of galaxies. Radio synchrotron emission provides the best
tools for measuring the strength and structure of galactic magnetic
fields \citep{beck96}. However, the sensitivity of present-day radio
observations only allows detailed studies for a couple of nearby
galaxies \citep{beck05}.

Rotation measures ($\RM$) towards polarized background sources
located behind spiral galaxies can trace regular magnetic fields in
these galaxies out to large distances, even where the synchrotron
emission of the galaxy itself is too weak to be detected
\citep{han98,gaensler05}. However, with the sensitivity of
present-day radio telescopes, the number density of polarized
background sources is only a few sources per solid angle of a square
degree, so that only M~31 and the LMC, the two angularly largest
galaxies in the sky, could be investigated so far
\citep{han98,gaensler05}.

Future high-sensitivity radio facilities will be able to resolve the
detailed magnetic field structure in galaxies by observing their
polarized intensity and $\RM$ directly. Furthermore, they will be
able to observe a huge number of faint radio sources, thus providing
the high density background of polarized point sources. This opens
the possibility of mapping the Faraday rotation of polarized
background sources towards nearby galaxies and study their magnetic
fields. The detection of large-scale field patterns (modes) and
their superpositions would strengthen dynamo theory for field
amplification and ordering, while the lack of $\RM$ patterns would
indicate a primordial origin or fields structured by gas flows
\citep{beck06}. Mapping the $\RM$ of the diffuse polarized emission
of galaxies at high frequencies is restricted to the star-forming
regions where cosmic-ray electrons are produced. At low frequencies
the extent of synchrotron emission is larger due to the propagation
of cosmic ray electrons, but Faraday depolarization increases much
faster with wavelength, so that the diffuse polarized emission is
weak. On the other hand, $\RM$ towards polarized background sources
are generated by regular fields plus ionized gas, which extend to
larger galactic radii, as indicated by the existing data sets
\citep{han98,gaensler05}, and are less affected by Faraday
depolarization.

A major step towards a better understanding of galactic magnetism
will be achieved by the Square Kilometre Array (SKA,
www.skatelescope.org) and its pathfinders, the Allen Telescope Array
(ATA) in the US, the Low Frequency Array (LOFAR) in Europe, the
MeerKAT in South Africa, and the Australia SKA Pathfinder (ASKAP).
The SKA will be a new-generation telescope with a square kilometre
collecting area, a frequency range of 70~MHz to 25~GHz with
continuous frequency coverage, a bandwidth of at least 25\%, a field
of view of at least 1 square degree at 1.4~GHz, and angular
resolution of better than 1~arcsecond at 1.4~GHz. The SKA is planned
to consist of three separate arrays: a phased array for low
frequencies (70--300~MHz), a phased array for medium frequencies
(300--1000~MHz), and an array of single-dish antennas for high
frequencies (1--25~GHz).

``Cosmic magnetism'' is one of six Key Science Projects for the SKA
with the plan to measure a grid of more than $10^7$ $\RM$ data
towards polarized sources over the whole sky \citep{gaensler04}.
This will allow measurement of the evolution of magnetic fields in
galaxies from the most distant to nearby galaxies and in the Milky
Way. Large-scale field structures should be observable via $\RM$
mapping of the diffuse polarized emission in galaxies
\citep{beck06}. Even more promising are $\RM$ measurements towards
polarized background sources, which are discussed in this paper.

%--------------------------------------------------------------------
\section{Models of Faraday rotation towards background sources in galaxies}
\label{sec:models}

A model for $\RM$ towards polarized background sources needs the
following ingredients: the distribution and sign of the regular and
random field of the galaxy, the distribution of electron density,
the density and fluxes of polarized background sources, and their
internal RM. The measured $\RM$ value towards each background source
is the sum of all rotation measure contributions along the line of
sight, including the Galactic foreground rotation and any internal
$\RM$ of the background source itself. The contribution of the
Galactic foreground has to be subtracted from the observed $\RM$
values before any further analysis, but this contribution is
presently known only for a coarse grid of background sources on
large scales of a few degrees \citep{kronberg81, johnston04}. $\RM$
data of background sources near the plane of the Milky Way reveal
fluctuations by several $100\FRM$ on scales of a few arcminutes
\citep{brown07}. The contribution of the Galactic $\RM$ can be
accounted properly from the density of $\RM$ detected with the LOFAR
or SKA in a slightly larger region surrounding each galaxy. The
internal $\RM$ of the background sources will average out if the
number of background sources used for the analysis is large enough.
Furthermore, a culling algorithm can be used to remove background
sources with high internal $\RM$ values. \cite{hennessy89} and
\cite{johnston04}, for example, removed sources with measured $\RM$
exceeding n-times the standard deviation from the mean (or median)
$\RM$.

\subsection{Models for the regular magnetic field in the disk}
\label{sec:diskfieldmodels}

The {\em mean-field $\alpha$--$\Omega$ dynamo model}\ is based on
differential rotation and the $\alpha$-effect \citep{beck96}.
Although the physics of dynamo action still faces theoretical
problems \citep[e.g.][]{brand05}, the dynamo is the only known
mechanism able to generate large-scale {\em coherent}\ (regular)
magnetic fields of spiral shape. These coherent fields can be
represented as a superposition (spectrum) of modes with different
azimuthal and vertical symmetries. In a smooth, axisymmetric gas
disk the strongest mode is the one with the azimuthal mode number
$m=0$ ({\em axisymmetric}\ spiral field), followed by the weaker
$m=1$ ({\em bisymmetric}\ spiral field), etc. \citep{elstner92}.
These modes cause typical variations in Faraday rotation along the
azimuthal direction in the galaxy disk \citep{krause90}. In flat,
uniform disks the axisymmetric mode with {\em even}\ vertical
symmetry (S0 mode) is excited most easily \citep{bary87}, while the
{\em odd}\ symmetry (A0 mode) dominates in spherical objects. The
timescale for building up a coherent field from a turbulent one is
$\approx 10^9$~yr \citep{beck94}.

Most galaxies reveal spiral patterns in their polarization vectors,
even flocculent or irregular galaxies. The field observed in
polarization can be anisotropic or regular. $\RM$ is a signature of
regular fields, while anisotropic fields generate polarized
emission, but no Faraday rotation. Large-scale $\RM$ patterns
observed in several galaxies \citep{krause90,beck05} show that at
least some fraction of the magnetic field in galaxies is regular.
The classical case is the strongly dominating axisymmetric field in
the Andromeda galaxy M~31 \citep{berk03,fletcher04}. A few more
cases of dominating axisymmetric fields are known \citep[e.g. the
LMC,][]{gaensler05}, while dominating bisymmetric fields are rare
\citep{krause89}. The two magnetic arms in NGC~6946 \citep{bh96},
with the field directed towards the galaxy's center in both, are a
signature of superposed $m=0$ and $m=2$ modes. However, for many of
the nearby galaxies for which multi-frequency observations are
available, angular resolutions and/or signal-to-noise ratios are
still too low to reveal dominating magnetic modes or their
superpositions. The regular field of the Milky Way reveals several
reversals in the plane, but the global structure is still unclear
\cite[][ Sun et al. in press]{han97,brown07}. The regular field near
the Sun is symmetric with respect to the Galactic plane.

For this paper, we restrict our analysis to the three lowest
azimuthal modes of the toroidal regular magnetic field: axisymmetric
spiral (ASS, $m=0$), bisymmetric spiral (BSS, $m=1$),
quadrisymmetric spiral (QSS, $m=2$), and the superpositions ASS+BSS
and ASS+QSS. All modes are assumed to be symmetric with respect to
the disk plane (S-modes). The maximum regular field strength is
assumed to be $5~\mu$G for all modes, consistent with typical values
from observations \citep{beck05}.

The radial extent of regular magnetic field is not known yet. In
NGC~6946, its scalelength is at least 16~kpc \citep{beck07}. $\RM$
measurements of polarized sources behind M~31 indicate that the
regular field strength out to at least 15~kpc radius is similar to
that in the inner disk \citep{han98}.

\subsection{Models for the magnetic field in the halo}
\label{sec:zfieldmodels}

Galactic dynamos operating in thin disks also generate poloidal
fields that extend far into the halo. They are about one order of
magnitude weaker than the toroidal fields in the disk \citep{ruz88}
and hence are irrelevant for our models. Polarization observations
of nearby galaxies seen edge-on generally show a magnetic field
parallel to the disk near the disk plane, but recent
high-sensitivity observations of several edge-on galaxies like M104,
NGC~891, NGC~253, and NGC~5775 show vertical field components that
increase with increasing height $z$ above and below the galactic
plane and also with increasing radius, so-called X-shaped magnetic
fields \citep[]{heesen05,heesen07,soida05,krause06,krause07}. Such a
vertical field can be due to a galactic wind that may transport the
disk field into the halo. Dynamo models including a wind outflow
show field structures quite similar to the observed ones
\citep{brand93}.

The detailed analysis of the highly inclined galaxy NGC~253 allowed
a separation of the observed field into an ASS disk field and a
vertical field \citep{heesen07}. The tilt angle between the vertical
field and the disk is about 45\degr. This value corresponds
approximately to the tilt angles at large $z$ observed in other
edge-on galaxies.

Assuming energy-density equipartition between magnetic fields and
cosmic rays, the scale height of the total magnetic field is about
4~times larger than the scale height of synchrotron emission of
typically about 1.8~kpc \citep{krause04}. As the degree of linear
polarization increases with height above the disk midplane, the
scale height of the regular field may be even larger.

\subsection{Thermal electron density models}
\label{sec:electronmodels}

In this paper we model the ionized gas of the disk by a Gaussian
distribution in the radial and vertical ($z$) directions. A scale
height of $1$\,kpc is adopted from the models of the free electron
distribution in the thick disk of the Milky Way fitted to pulsar
dispersion measures \citep{gomez01,cordes02}. The electron density
near the galaxy center is assumed to be 0.03~cm$^{-3}$, the same as
in the Milky Way models.

The radial scalelength of the density $n_e$ of free electrons in the
Milky Way of $\approx$15~kpc \citep{gomez01} is much larger than the
scalelength of radio thermal emission in nearby galaxies, e.g. about
4~kpc in NGC~6946 \citep{walsh02}, 3~kpc in M~33 \citep{taba07}, and
5~kpc for the H$\alpha$ disk of the edge-on galaxy NGC~253 (Heesen,
priv. comm.). As the thermal emission scales with $n_e^2$, the
scalelength of $n_e$ is twice larger, but still smaller than what is
quoted for the Milky Way. As a compromise, we use a Gaussian
scalelength of 10~kpc in our model.

\begin{figure}
\begin{center}
  \includegraphics[width=60mm]{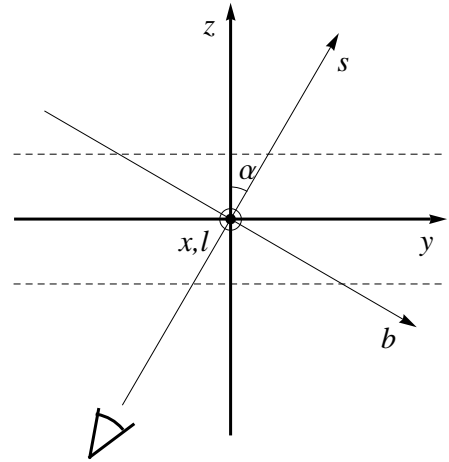}
  \caption{The two coordinate systems used in this paper:
  $l,b$ in the sky plane and $s$ along the line of sight, and
  $x,y,z$ in the model galaxy plane and perpendicular to it. The dashed lines
  indicate the thickness of the disk.}
  \label{fig:coord}
\end{center}
\end{figure}

The distribution of ionized gas follows that of the optical spiral
arms. The arm-interarm-contrast of the H$\alpha$ emission is between
2 and 10 in typical spiral galaxies like NGC~6946 \citep[see the
smoothed H$\alpha$ map shown in][]{frick01}. On the other hand,
regular magnetic fields observed in polarization are strongest in
interarm regions \citep{bh96,frick01} with a similar
arm-interarm-contrast. As Faraday rotation in NGC~6946 is also
highest in the interarm regions \citep{beck07}, the azimuthal
variation in Faraday rotation, determined by the product of regular
field strength and electron density, is mostly determined by the
azimuthal variation of the regular field. We conclude that the
azimuthal variation in electron density can be neglected in our
models.

In summary, we adopt a Gaussian dependence of $n_e$ on galactic
radius $r$ (in the galactic plane) and on height $z$ above the disk
midplane:

\begin{equation}\label{modelne}
 n_e(r,z) = n_0 \exp{\left\{-\left(\frac{z}{h}\right)^2\right\}}
   \exp{\left\{-\left( \frac{r}{r_0}\right)^2\right\}},
\end{equation}
where $n_0 $=0.03~cm$^{-3}$ is the electron density near the galaxy
center, $r_0$=10~kpc is the radial scalelength of the $n_e$
distribution in the galaxy model, and $h$=1~kpc is its vertical
scalelength.

\subsection{Faraday rotation models}
\label{sec:rmmodels}

Two Cartesian coordinate systems are used: the coordinate system
$l,b,s$ in the sky plane with the same orientation as the galaxy
($l,b$ are the distances counted from the center of the observed
galaxy along the major and minor axes of the projected disk, and $s$
corresponds to the direction of the line of sight) and the system
$x,y,z$ attached to the galaxy ($x,y$ are in the galactic midplane
and $z$ is perpendicular to it). The coordinates $x$ and $l$ of both
systems coincide and $i$ is the angle between coordinates $z$ and
$s$ (or $y$ and $b$, see Fig.~\ref{fig:coord}), called the
inclination angle of the galactic plane (where $i=0^{\circ}$ means
face-on).

\begin{figure*}
\begin{center}
\includegraphics[width=0.8\textwidth]{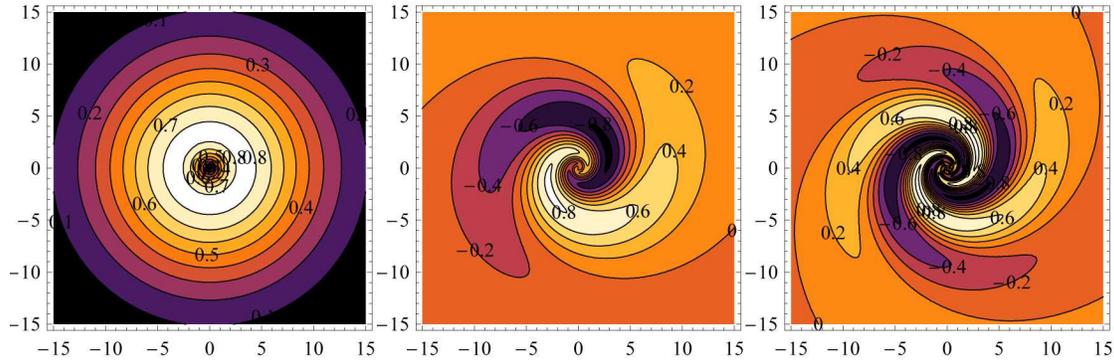}
  \caption{Modeled strength of the {\it regular} magnetic field multiplied by
  the thermal electron density in the galactic midplane (i.e. in the
  plane $x,y$) for the field modes $m=0$ (ASS), $m=1$ (BSS), and $m=2$ (QSS) for
  $B=1$.}
  \label{fig:bss1}
\end{center}
\end{figure*}

The galactic magnetic field is modeled as a superposition of a
regular part with a simple azimuthal symmetry $B_m$ and a random
part $B_{turb}$, which describes the contribution of large-scale
galactic turbulence:
\begin{equation}
\vec B(\vec r) = \vec B_m(\vec r) + \vec B_{turb}(\vec r),
\end{equation}
where $\vec r$ is the radius vector in galactic coordinates, which
will be written in Cartesian coordinates (x,y,z) or cylindrical
ones ($r=\sqrt{x^2+y^2}$, $\phi=\arctan{y/x}$, $z$).

First, the regular part of magnetic field is supposed to be purely
horizontal ($B_z=0$), parameterized by the pitch angle $p$ and the
intensity $B_m$ of the corresponding azimuthal mode $m$ (we only
consider $m=0,1,2$, using the notation ASS, BSS, and QSS, see
Fig.~\ref{fig:bss1}). The BSS mode has two reversals along azimuthal
angle, the QSS mode has four reversals. All modes are symmetric with
respect to the disk plane. The strength of the magnetic field in
cylindrical coordinates is defined as
\begin{equation}\label{model0}
B_m (r,\phi,p) =
 B \cos{\left( m \left(\frac{ \ln{r}}{\tan p}  -
 \phi+\phi_0\right)\right)}
  \tanh{\left(\frac{5r}{r_0}\right)},
\end{equation}
where $B$ is the field amplitude (strength) and $\phi_0$ the
azimuthal phase of the mode. The $\tanh$ term is introduced to
suppress the field near the center of the galaxy where the spiral
field is strongly twisted. Equation (\ref{model0}) does not include
any decrease in the field strength with radius $r$ and height $z$,
because the decay in regular field strength is much slower than that
of electron density (see Eq.(\ref{modelne})).

The $\RM$ is proportional to the product of the density of thermal
electrons $n_e$ and the component of the regular magnetic field
$\Brpa$ parallel to the line of sight. Note that we cannot separate
the contributions of $B$ and $n_e$ using the $\RM$ data alone and
that we can reconstruct only the product ($\Brpa \, n_e$). As we
neglect any azimuthal (e.g. spiral) structure in $n_e$, the
small-scale structure in $\RM$ is determined by $B$, whereas the
general radial and vertical decrease in $\RM$ is determined by
$n_e$. The distribution of this field in the galactic midplane for
$m=0,1,2$ (ASS, BSS, and QSS) is shown in Fig.~\ref{fig:bss1}.

\begin{figure}
\begin{center}
\includegraphics[width=60mm]{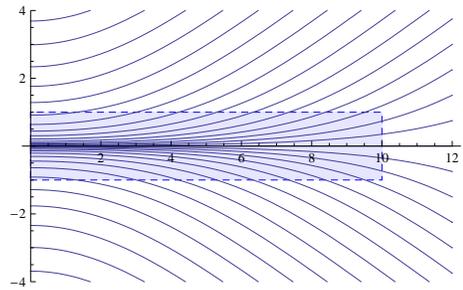}
  \caption{Orientation of the halo magnetic field.
  The box sketches the region of ionized gas $r <r_0$, $-h <z <h$.}
  \label{fig:tilt}
\end{center}
\end{figure}

Second, we consider a model in which a vertical field $B_z$ exists
and increases with height $z$, according to the magnetic field
structure observed in the halos of nearby edge-on galaxies (see
Sect.\, \ref{sec:zfieldmodels}). We suppose that the modulus of the
magnetic field follows again Eq.~(\ref{model0}), but the field lines
are tilted with respect to the plane. The tilt angle $\chi$ is
defined as
\begin{equation}\label{chi}
\chi = \chi_0 \tanh{\frac{z}{2h}} \tanh{\frac{3r}{r_0}},
\end{equation}
where $\chi_0 = \pi/4$ is the limit of the tilt angle in our model
achieved at $r\approx r_0/3$, $z\approx 2h$. The orientations of
magnetic field lines for the ASS model are shown in
Fig.~\ref{fig:tilt}.

The magnetic field component $\Brpa$ parallel to the line of sight
includes the turbulent (random) magnetic field as well as the
regular one, so that

\begin{equation}
\label{RM} \RM(l,b) = 0.81 \int_{-\infty}^{\infty}  \left( \vec
B_m(\vec r,p)+\vec B_{turb}(\vec r)  \right) n_e(\vec r)  d \vec
s.
\end{equation}
The turbulent part of magnetic field $B_{turb}$ should describe the
three-dimensional random field with given spectral properties in the
whole range of scales. To avoid full 3-D simulations of the random
vector field and its contribution to $\RM$, we prefer to model this
contribution of the irregular part of magnetic field directly in the
$\RM$ maps. Thus we write

\begin{equation}
\label{RMr} \RM(l,b) = 0.81 \int_{-\infty}^{\infty}   \vec
B_m(\vec r,p)n_e(\vec r) d \vec s  + \RM_{turb}(l,b),
\end{equation}
where we add the random part of the galactic field projected to the
sky plane, i.e. integrated along the line of sight.

\begin{figure*}
\begin{center}
\includegraphics[width=0.8\textwidth]{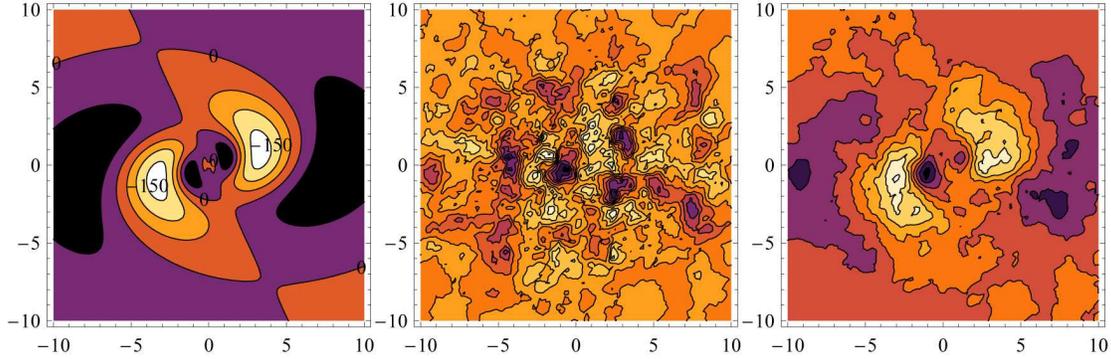}
\caption{Modeled $\RM$ maps (in $\FRM$) for an inclination angle of
$i=10^o$ generated for a pure BSS field pattern, for $B=5\,\mu$G and
$n_0 $=0.03~cm$^{-3}$ (left), a random turbulent field $\RM_{turb}$
for $\Delta\RM_{turb}=30~\FRM$ (middle), and their superposition
(right).} \label{fig:rmr}
\end{center}
\end{figure*}

$\RM_{turb}(l,b)$ is produced in a way that allows for the spectral
and spatial distribution of galactic turbulent fields. \cite{vogt05}
show that by observing a turbulent magnetic field with a
three-dimensional Fourier power-law spectrum like $|\hat{\vec
B}(\vec k)|^2 \sim k^\alpha$ one obtains an $\RM$ map, the {\it
two-dimensional} spectrum, from which another power law
follows $|\hat{\RM} (\vec{k_\perp})|^2 \sim {k_\perp}^\beta$ with
the slope $\beta = \alpha$. Here the hat is used for the notation of
the Fourier transform of any function $\hat f(\vec k) = \int f(\vec
x) e^{i\vec k \vec x} d \vec x$, $\vec k$ is the 3-D wave vector and
$\vec{k_\perp}$ is the 2-D wave vector, defined in the plane of the
$\RM$ map. It is supposed that the magnetic spectrum is
Kolmogorov-like ($E_B(k) \sim k^2 |\hat{\vec B}(\vec k)|^2 \sim
k^{-5/3}$, hence $\alpha = -11/3$) at large wave numbers (on scales
smaller than the thickness of the galactic disk) and is growing at
small wave numbers as $E_B(k) \sim k^2$ (then $\alpha=0$). Thus we
generate a random map in the Fourier space with the required
spectrum of the form:
\begin{eqnarray}
\label{rmfourier}
 |\hat{\Phi}(\vec k_\perp)|^2 =
\left\{
  \begin{array}{ll}
   \Phi_0, & k<k_0 \\
     \Phi_0\left( k/k_0 \right)^{-11/3}, & k>k_0.
  \end{array}
\right.
\end{eqnarray}
Here $\Phi_0$ defines the amplitude of the turbulent part of the
generated map, and $k_0$ is the wave number corresponding to the
turbulence macroscale ($k_0=5$ is adopted here to reach a realistic
spectrum with a maximum on scales corresponding to the typical size
of supernova remnants). Hereafter we return to physical space and
obtain the map $\RM_{turb}(l,b)$ as the product of the random map
$\Phi(l,b)$ and the galactic ``electron thickness'':

\begin{equation}\label{rmturb}
 \RM_{turb}(l,b) = \Phi(l,b) \, DM(l,b).
\end{equation}
Namely, $DM(l,b)$ is the dispersion measure for the given line of
sight:

\begin{equation}\label{sne}
 DM(l,b) = \int_{-\infty}^{\infty} n_e(\vec r) d \vec s.
\end{equation}
The amplitude of the signal in this map is normalized by the
parameter $\Phi_0$, so that the r.m.s. $\Delta\RM_{turb}$ of the
random field $\RM_{turb}$ adopts some given value. We will consider
three cases: weak turbulence ($\Delta\RM_{turb}=10~\FRM$), moderate
turbulence ($\Delta\RM_{turb}=30~\FRM$), and strong turbulence
($\Delta\RM_{turb}=50~\FRM$). Note that, due to the normalization,
$\Delta \RM_{turb}$ does not depend on the amplitude of $DM$ and
hence does not increase with increasing inclination, although such a
behavior may occur in real galaxies (see Sect.~\ref{sec:disc}).

Figure \ref{fig:rmr} shows an example of the $\RM$ map calculated
for a galactic field observed almost face-on (inclination angle
$i=10^\circ$), simulated for a pure BSS field pattern (left), the
map of turbulent field $\RM_{turb}$, and the resulting $\RM$ map for
the same BSS model superimposed by a moderate level of turbulence
with a ``realistic'' spectrum (right).

%--------------------------------------------------------------------
\section{Number density of polarized background sources}
\label{sec:sources}

Future high-sensitivity radio facilities (LOFAR, SKA) will be able
to observe faint radio sources, thus providing a high-density
background of polarized sources. This opens a new possibility of
mapping the Faraday rotation of polarized background sources towards
nearby galaxies and studying their magnetic fields. In this section
we estimate the density of the polarized sources towards a typical
spiral galaxy placed at different distances from the observer and
study the errors associated with rotation measures and their
dependence on the telescope specifications.

\begin{figure}
\begin{center}
%\resizebox{\hsize}{!}{\includegraphics[angle=-90]{Nc-P.ps}}
\resizebox{\hsize}{!}{\includegraphics[angle=-90]{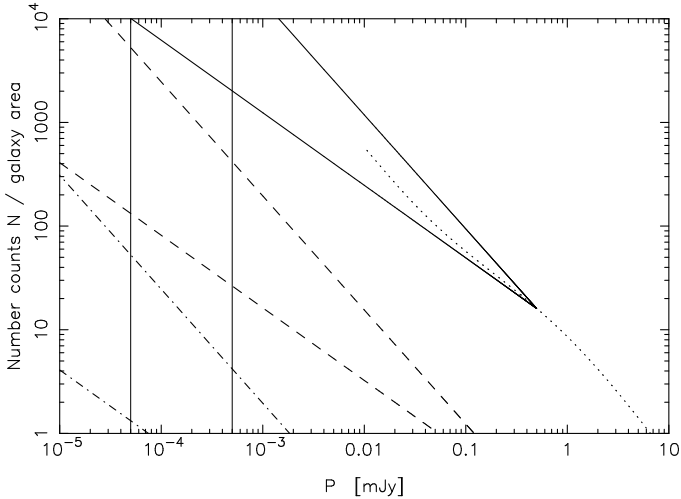}}
  \caption{The dotted line shows the cumulative count $N$ of
  polarized sources brighter than the polarized flux density $P$ at $\lambda$=21~cm,
  taken from \cite{taylor07}, extrapolated to 0.01~mJy polarized flux density
  and normalized to source numbers per 1~deg$^2$ solid
  angle on sky. The possible range of extrapolated number counts per galaxy solid
  angles of 1~deg$^2$ are shown as thick solid lines, 0.013~deg$^2$ as dashed
  lines,
  and 0.00013~deg$^2$ as dot-dashed lines. These solid angles correspond to
  distances of a galaxy (with $r_0=10$\,kpc) at 1.15\,Mpc, 10\,Mpc, and
  100\,Mpc. The two different lines at each distance refer to slopes
  $\gamma$ of the extrapolated source numbers of $0.7$ (lower line)
  and $1.1$ (upper line). The two thin vertical lines represent the
  $5\sigma_p$ detection limit for sources observed at $\lambda$=21~cm with the
  SKA within 100 hours (left) and 1 hour (right) observation time.}
  \label{fig:ncp}
\end{center}
\end{figure}

The differential number counts of polarized sources has been
recently studied using the deep polarimetric observations at 1.4~GHz
of the European Large Area \emph{ISO} Survey North 1 region (ELAIS
N1) as a part of the DRAO \emph{Planck} Deep Fields project
\citep{taylor07}. They show that the observed distribution of number
counts in differential representation (i.e. numbers per flux density
interval) is almost flat down to polarized flux densities of
$\approx$~0.5~mJy. They also show that the extrapolated differential
number counts of polarized sources decreases down to the 0.01~mJy
level. The corresponding cumulative source number counts is shown in
Fig.~\ref{fig:ncp}, normalized to the solid angle of 1~deg$^2$,
which roughly corresponds to the area covered by the inner part of
the nearest galaxy M~31 \citep{han98}.

Observations with the SKA (assuming 10$^6$~m$^2$ collecting area,
50~K receiver system temperature and $\Delta\nu$/$\nu =
\Delta\lambda$/$\lambda=0.25$ relative bandwidth) will achieve an
r.m.s. noise of 0.1~$\mu$Jy/beam area and 0.01~$\mu$Jy/beam area at
1.4~GHz ($\lambda$=21~cm) within one hour and 100 hour integration
time, respectively \citep{carilli04}.

We extrapolate the source counts at $\lambda$=21~cm from
$P_0=0.5$~mJy to the limiting flux density $P_{min}$ of
0.05~$\mu$Jy, assuming a power-law relation between the cumulative
number counts per deg$^2$ ($N_{\square}$) and polarized flux density
($P$):
\begin{equation}
 N_{\square}(>P)= N_0 \left( \frac{P}{P_0}\right)^{-\gamma},
 \label{a2}
\end{equation}
where $N_0=16$~deg$^{-2}$ is the cumulative source number at
$P=P_0=0.5$~mJy. The exponent $\gamma$ is expected in the range
between 0.7 and 1.1 (``pessimistic'' and ``optimistic'' number
counts). The {\em number count per solid angle} of $2 r_0 \times 2
r_0$ around a typical face-on spiral galaxy with a radius of $r_0$
at a distance $D$ is
\begin{equation}
N=N_0\Omega_g \left( \frac{P}{P_0}\right)^{-\gamma}, \label{a}
\end{equation}
where $\Omega_g = (2r_0/D)^2(180/\pi)^2$ is the solid angle of the
galaxy, given in square degrees.

In Fig.~\ref{fig:ncp} we show the predicted number counts per
typical spiral galaxy ($r_0=10$\,kpc, $20\times20$\,kpc$^2$ area)
observed at a distance of 1.15\,Mpc (1\,deg$^2$ solid angle),
10\,Mpc (0.013\,deg$^2$) and 100\,Mpc (0.00013\,deg$^2$). With the
``quick-look'' sensitivity of 0.01~mJy, the SKA will already detect
within a few minutes about 1000 polarized sources towards the
nearest spiral galaxy M~31. With the best achievable SKA sensitivity
of $P\approx$~0.05~$\mu$Jy, about 5$\times$10$^4$ polarized sources
behind M~31 and hundreds to tens of polarized background sources
will be detected behind a typical galaxy at a distance between 10
and 100~Mpc. Only for the pessimistic case ($\gamma=0.7$), the
source number becomes very small for galaxies near 100~Mpc distance.

The instrumental 1$\sigma$ error in rotation measure $\Delta \RM$ of
each simulated radio source of polarized flux density ($P$) in the
cumulative number counts is
\begin{equation}
\label{rma}
  \Delta \RM = \frac{\sigma_\phi}{2\,\lambda^2 \, \frac{\Delta\lambda}{\lambda}},
\end{equation}
where $\sigma_\phi$ is the 1$\sigma$ error (in rad) of the total
difference in polarization angles $\phi$ within the observed
wavelength range, $\lambda$ the center wavelength, and
$\Delta\lambda/\lambda$ the relative bandwidth of the observations.
If the instrumental r.m.s. noise in polarized flux density is almost
constant across the whole band, $\sigma_\chi$ is determined by the
errors of polarization angles in the frequency channels at the
smallest and largest wavelength, which for simplicity are assumed to
be the same. Then,
\begin{equation}
\label{rmb}
  \sigma_\phi = \sqrt{2}\,\frac{\sigma_p}{2\,P},
\end{equation}
where $\sigma_p$ is the instrumental r.m.s. noise in polarized flux
density per frequency channel and $P$ is the polarized flux density
of the source. Hence,
\begin{equation}\label{rmc}
  \Delta \RM = \frac{A}{P},
\end{equation}
where
\begin{equation}
\label{a1}
  A = \frac{\sigma_p}{2\,\sqrt{2}\,\lambda^2 \, \frac{\Delta\lambda}{\lambda}}.
\end{equation}
The ``sensitivity parameter'' $A$ only depends on instrumental
specifications. It becomes smaller for lower noise $\sigma_p$
(larger integration time of the telescope), larger relative
bandwidth, and longer observation wavelength. For the expected
performance of the SKA at $\lambda$=21~cm (see above), A~$\approx
3~\mu$Jy~rad~m$^{-2}/\sqrt{t}$ where $t$ is the observation time in
hours. For a more general discussion of $\RM$ errors, see Appendix A
of \cite{brentjens05}.

The maximum error of $\RM$ is reached at the limiting flux density
$P_{min}$ of the survey. For $P_{min}\approx 5 \sigma_p$ one derives
\begin{equation}\label{rm_max}
  \Delta \RM_{max}(P_{min}) = \frac{A}{P_{min}} \approx \frac{1}{14
  \lambda^2 \, \frac{\Delta \lambda}{\lambda}},
\end{equation}
which is independent of noise $\sigma_p$ and of observation time.
For the SKA $\Delta \RM_{max}$ is $\approx 6$~rad~m$^{-2}$.
For comparison, the VLA (using two bands centered at $\lambda$=22~cm
and 18~cm, see Sect.~\ref{sec:m31}) gives $\Delta \RM_{max} \approx
10$~rad~m$^{-2}$.

Typical maximum values of $\RM$ are $\RM~\approx$~40~rad~m$^{-2}$
for a galaxy inclined by $i=10^\circ$ (almost face-on) and
$\RM~\approx$~200~rad~m$^{-2}$ for $i=90^\circ$ inclination
(edge-on). With the SKA centered at $\lambda$=21~cm, the maximum
error $\Delta \RM_{max}$ is significantly smaller than the typical
maximum values of $\RM$ for edge-on and face-on nearby galaxies.

\begin{figure}
\begin{center}
%\resizebox{\hsize}{!}{\includegraphics[angle=-90]{sigP-D.ps}}
\resizebox{\hsize}{!}{\includegraphics[angle=-90]{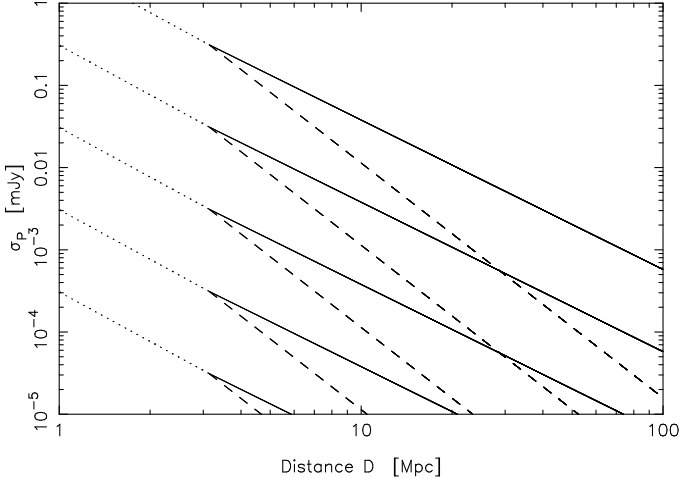}}
  \caption{Relation between instrumental noise and distance to the
  galaxy (see Eq.~(\ref{q})) for different values of the ``observation
  parameter'' $Q=0.001, 0.01, 0.1, 1, 10$~rad~m$^{-2}$ (from left to
  right) drawn for pessimistic and optimistic slopes, $\gamma=0.7$
  (dashed line) and $\gamma=1.1$ (solid line) for $\lambda=21$\,cm,
  $\delta\lambda/\lambda = 0.25$. The dotted line corresponds to the
  observed slope $\gamma\approx 1$ (see Fig.~\ref{fig:ncp}). The numbers
  of sources per solid angle of the galaxy corresponding to the above $Q$
  values are $N_{max}=462, 92, 18, 4, 1$ for $\gamma=0.7$ and
  $N_{max}=15411, 1224, 97, 8, 1$ for $\gamma=1.1$ (Eq.~(\ref{nmax})).}
  \label{fig:sigP}
\end{center}
\end{figure}

From Eqs.~(\ref{rmc}) and (\ref{a2}--\ref{a}) one may recover
\begin{equation}\label{drmq}
  \Delta\RM = Q\,N^{1/\gamma},
\end{equation}
where
\begin{equation}\label{q}
  Q =  A\,(N_0\Omega_g)^{-1/\gamma} P_0^{-1}.
\end{equation}
The {\em ``observation parameter''} $Q$ only depends on
observational characteristics. It defines $\Delta \RM$ for the
brightest source statistically expected within the solid angle of a
given galaxy by a given telescope (i.e. for N=1 in
Eq.~(\ref{drmq})). It varies with noise $\sigma_p$ and with distance
$D$ as $\sigma_p D^{2/\gamma}$. The value of $Q$ lies in the range
between $\approx 0.001$ rad m$^{-2}$ and $\approx 10$ rad m$^{-2}$
for $\sigma_p $ between $0.01\,\mu$Jy and 0.03\,mJy and the range of
distances from 1\,Mpc to 100\,Mpc (Fig.~\ref{fig:sigP}). From
Eqs.~(\ref{a}) and (\ref{q}), the parameter $Q$ directly relates to
the maximum number of sources $N_{max}$ (above the $5\sigma_p$
detection limit $P_{min}$) per solid angle of the galaxy:

\begin{equation}\label{nmax}
  N_{max}(P_{min}) = \left(10 \sqrt{2}\, \lambda^2
  \frac{\Delta\lambda}{\lambda}\right)^{-\gamma} Q^{-\gamma}.
\end{equation}

%--------------------------------------------------------------------

%--------------------------------------------------------------------
\section{Recognition of regular galactic field structures from a
limited sample of background sources} \label{sec:fitting}

In this section we describe and use the $\chi^2$ minimization method
to test the magnetic field models in the disk and halo of a galaxy.
Application of this method to recognize the magnetic field structure
in M\,31 is also presented.

\subsection{Method}

Suppose that $N$ polarized background sources are observed with an
``observation parameter'' $Q$ towards a galaxy with radius $r_0$ at
distance $D$. Given the Faraday rotation values of the model
(${\rm{RM}}_{mod}$) caused by the regular field (${\rm{RM}}_{reg}$)
and the turbulent field in the galaxy, we want to recognize the
structure of the regular magnetic field, such as ASS, BSS, QSS, or
their combinations, and determine the best-fit parameters of the
model (amplitude, spiral pitch angle, and azimuthal phase) and their
standard deviations.

\begin{figure}
%\centerline{\includegraphics[width=0.35\textwidth]{Neff.eps}}
\centerline{\includegraphics[width=0.35\textwidth]{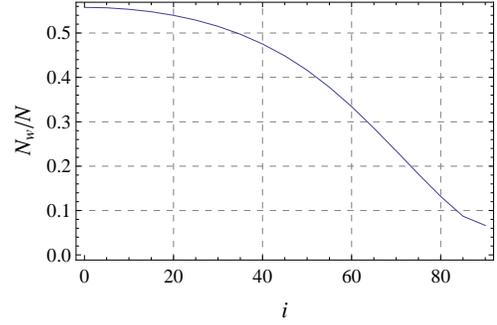}}
 \caption{Factor of effective points vs inclination angle $i$ due to
 geometry of the disk of thermal electrons}.
 \label{figNeff}
\end{figure}

For a fixed inclination angle of the galaxy, we simulate a
``template'' of Faraday rotation values
$\RM_{reg,\,n}(\xi_1,...,\xi_d)$ of $N$ points (with coordinates
$(l_n,b_n)$) for the regular field only where $\xi_j$ are the free
parameters of the magnetic field model: the amplitude and spiral
pitch angle of the magnetic field (number of free parameters $d=2$
for ASS), the azimuthal phase ($d=3$ for BSS and QSS), and the
relative amplitude $q$ for the ASS field in combination with the BSS
or QSS field where the BSS or QSS field has the amplitude of $1-q$
($d=4$ for the combination). For the same set of points we simulate
the ``observed'' $\RM_{mod,\,n}$ (Eq.~(\ref{RM})) which takes both
the regular and turbulent magnetic fields into account. By varying
the free parameters we minimize the normalized $\chi^2$ value
between the ``observed'' and ``template'' $\RM$ maps,

\begin{equation}
\chi^2 = \frac{1}{N_w-d} \sum_{n=1}^N \frac{(\RM_{mod,\,n} -
\RM_{reg,\,n}(\xi_1,...,\xi_d))^2 w_n }{(\Delta\RM_{noise,n}^2 +
\Delta\RM_{turb}^2w_n^2)} \label{chi2},
\end{equation}
and estimate the best-fit parameters for each model. ($\Delta
\RM_{noise}$ and $\Delta\RM_{turb}$ are the instrumental noise
(Eq.~(\ref{drmq})) and the noise of the random field, respectively.)
The coefficients $w_n$ define the weight of each point $w_n =
DM(l_n,b_n)/DM(0,0)$, hence the contribution of each point to the
minimization procedure. The weight of the point $w_n$ is defined by
the dispersion measure $DM(l_n,b_n)$ (Eq.~(\ref{sne})) along the
line of the sight. Introducing this weight is needed to balance the
contribution of strong $\RM$ sources with small error, which are
located at the galactic periphery, where the low thermal electron
density does not allow to obtain reliable information about the
galactic magnetic field. The total value of $w_n$ gives the weighted
(effective) number of points $N_{w}=\sum_n w_n$. This number is
different from the real number of points $N$ taken per $2 r_0\times
2 r_0$ area. It also depends on the inclination angle $i$ (see
Fig.~\ref{figNeff}). This dependence is similar to $\cos i$, but it
is not equal to 1 at $i=0^\circ$ because of the Gaussian
distribution of the thermal electrons and not equal to $0$ at
$i=90^\circ$ due to the finite thickness of the disk. The
inclination angle $45^\circ$ yields $N_{w}=0.44 N$. This means that
in order to obtain at least $N_w=4$ we need a minimum of $N=10$.

This method is suited for simple field structures and a relatively
small number of $\RM$ points allowing the limited number of
parameters to be fitted. In the case of complicated field structures
and/or a large number of $\RM$ points, a full reconstruction should
be attempted (Sect.~\ref{sec:deconv}).

\subsection{Fitting the simulated data}
 \label{sec:results}

\begin{figure}
\centerline{\includegraphics[width=0.5\textwidth]{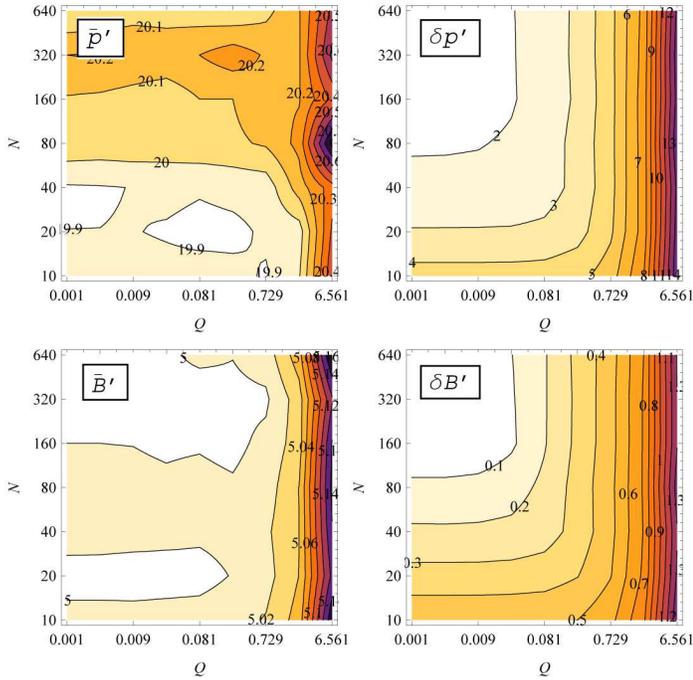}}
%\centerline{\includegraphics[width=0.5\textwidth]{8678fig8.eps}}
 \caption{ASS field model: mean values (figures in left column) and
 standard deviations (figures in right column) of pitch angles
 $\bar{p}'$ and amplitudes $\bar{B'}$ (given in $\mu$G) for fixed
 values of $h=1\kpc$, $p=20^{\circ}$, $B=5\,\mu$G, $i=45^{\circ}$ and
 assuming the pessimistic case of number counts ($\gamma=0.7$).
 $N$ is the real number of points observed within the solid angle of
 the galaxy and $Q$ is the observation parameter (Eq.~(\ref{q})).}
 \label{figc1}
\end{figure}

Throughout this paper we use typical values of $r_0=10$\,kpc,
$h=1\kpc$, $p=20^{\circ}$, $\n=0.03\cmcube$ in Eq.~(\ref{modelne})
and $B=5\,\mu$G in Eq.~(\ref{model0}). We start with recognition of
an ASS field as the simplest model. For a fixed $i=45^{\circ}$, we
simulate the $\RM$ values, noise distribution $\Delta \RM$, and
turbulent components of $N$ uniformly distributed background
sources. We determine the best-fit parameters $p'$ and $B'$ by
applying the $\chi^2$ minimization technique (Eq.~(\ref{chi2})).
Repeating this procedure for several thousand realizations we
calculate the mean pitch angle ($\bar{p}'$), amplitude ($\bar{B'}$),
and their standard deviations ($\delta p'$ and $\delta B'$). The
dependence of the statistical properties of the fitted parameters on
the number $N$ of sources per solid angle of the galaxy and on $Q$
are shown in Fig.~\ref{figc1}.

The simulated value of $\bar{p}'$ is in good agreement with the
value $p=20^{\circ}$ used for the model calculation over a wide
range of $N$ and $Q$. However, the standard deviation of $\delta p'$
is quite small only for low values of $Q$ and for large numbers $N$.
The amplitude $\bar{B'}$ is well defined, too. The behaviour of
$\delta B'$ is the same as for $\delta p'$. The standard deviation
diagrams show that there is a certain value of $Q$ below which
better observations do not improve the result of fitting. The same
is valid beyond a certain value of $N$. For example, for the level
of $\delta p'=3^{\circ}$ no more than 36 sources observed with
$Q=0.2$ are needed, or no better values than $Q=0.05$ when observing
21 sources. Hence, the fitting problem can be characterized by two
asymptotic values $N^*$ and $Q^*$, provided that the accuracy of the
fitting is fixed. (For the given example one obtains $N^*=21$ and
$Q^*=0.2$.) These two parameters define the required quality of
observations.

\begin{figure}
%\centerline{\includegraphics[width=0.4\textwidth]{resASS2.eps}}
\centerline{\includegraphics[width=0.4\textwidth]{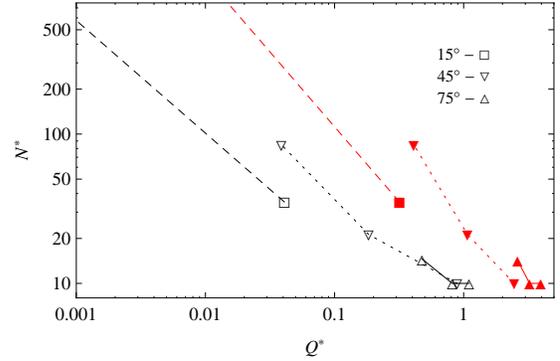}}
 \caption{Required asymptotic values
 $N^*$ and $Q^*$ to recognize an ASS-type field.
 The points correspond to $\delta p'=3^{\circ}$ for three different levels
 of turbulence (from bottom to top in each triplet: 10~$\FRM$, 30~$\FRM$ and
 50~$\FRM$) and for different inclination angles ($15^{\circ}$, $45^{\circ}$,
 $75^{\circ}$) for fixed $h=1\kpc$, $p=20^{\circ}$. Filled symbols correspond
 to the optimistic case ($\gamma=1.1$), open symbols to the pessimistic one
 ($\gamma=0.7$).}
 \label{figc2}
\end{figure}

In Fig.~\ref{figc2} we show the dependence of asymptotic parameters
$N^*$ and $Q^*$ on the level of turbulence and pitch angle of the
observed galaxy for optimistic and pessimistic values of $\gamma$.
The required level of accuracy is fixed at $\delta p'=3^{\circ}$.

We conclude that, in the case of the pessimistic slope, one needs
observations with smaller $Q$, hence smaller noise $\sigma_p$, which
requires a longer integration time. An increase in the number of
sources (for a fixed noise value $\sigma_p$) does not help. This can
be explained by the fact that for low values of $\gamma$ additional
sources are fainter, have higher errors, and do not improve the
result. The optimistic evaluation of $\gamma$ essentially increases
the values of $Q^*$, but slightly affects the values of $N^*$. As
expected, the result of fitting strongly depends on the level of
turbulence of the galactic field: in the case of intense turbulence
one needs more points {\em and} smaller $Q$ (longer observations or
a more sensitive instrument). The dependence on turbulence becomes
dramatic for weakly inclined (face-on) galaxies -- the reliable
fitting requires a huge number of sources under any value of
$\gamma$.

\begin{figure}
\centerline{\includegraphics[width=0.5\textwidth]{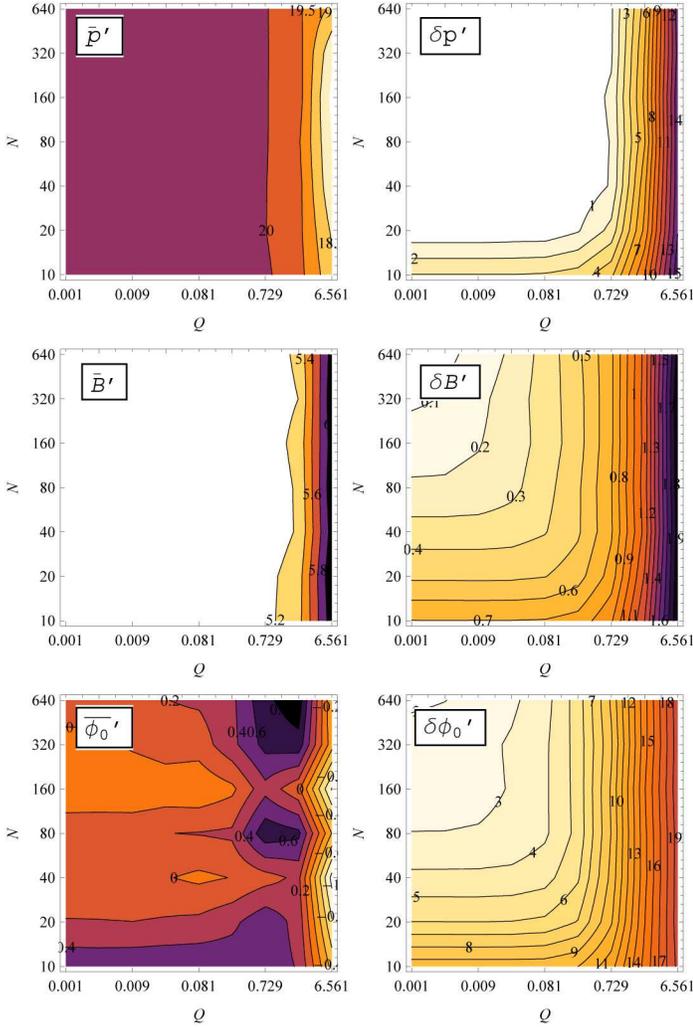}}
%\centerline{\includegraphics[width=0.5\textwidth]{8678fig10.eps}}
 \caption{BSS field model: Mean values (figures in left column) and standard
 deviations (figures in right column) of fitted pitch angles $\bar{p}'$,
 amplitudes $\bar{B'}$, and azimuthal phases $\bar{\phi_0}'$ for fixed
 $h=1\kpc$, $p=20^{\circ}$, $\phi_0=0^\circ$, and $i=45^{\circ}$, and
 assuming the pessimistic case of number counts ($\gamma=0.7$).
 $N$ is the real number of points observed within the solid angle
 of the galaxy and $Q$ the observation parameter.}
 \label{figc3}
\end{figure}

\begin{figure}
%\centerline{\includegraphics[width=0.4\textwidth]{resBSS2.eps}}
\centerline{\includegraphics[width=0.4\textwidth]{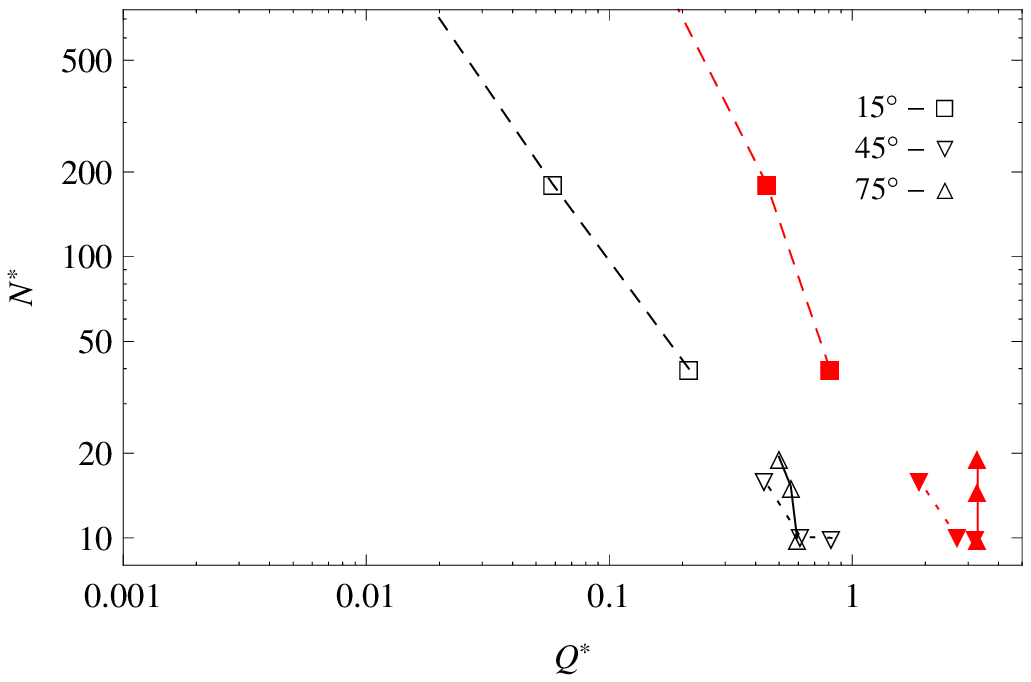}}
 \caption{Required asymptotical values $N^*$ and $Q^*$ to recognize a BSS-type field.
 The points correspond to $\delta p'=3^{\circ}$ for three different levels
 of turbulence (from bottom to top in each triplet: 10~$\FRM$, 30~$\FRM$ and
 50~$\FRM$) and different inclination angles ($15^{\circ}$, $45^{\circ}$,
 $75^{\circ}$) for fixed $h=1\kpc$, $p=20^{\circ}$. Filled symbols correspond
 to optimistic case ($\gamma=1.1$), open symbols to the pessimistic one
 ($\gamma=0.7$).}
 \label{figc4}
\end{figure}

\begin{figure}
%\centerline{\includegraphics[width=0.5\textwidth]{resAssBssProp.eps}}
\centerline{\includegraphics[width=0.5\textwidth]{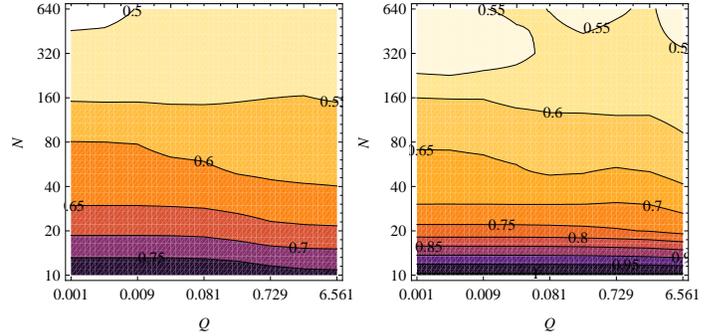}}
 \caption{Probabilities calculated from $\chi^2$ for ASS (left) and BSS (right).}
 \label{figProp}
\end{figure}

The same simulations are performed for the BSS and QSS models. An
additional free parameter (the phase $\phi$) is accounted for, but,
in general, the results are similar to those obtained for the ASS
model (see Figs.~\ref{figc3} and \ref{figc4}). Comparing
Figs.~\ref{figc2} and \ref{figc4} (or Figs.~\ref{figc1} and
\ref{figc3}), we conclude that the bisymmetric field can be easier
recognized (it requires fewer points and higher $Q$ for the same
accuracy of pitch angle definition) and weakly depends on the
turbulent component of magnetic field. The BSS field has better
chances of being recognized in slightly inclined galaxies. However,
with values of $N$, the BSS model has a higher probability of being
false than the ASS model (see Fig.~\ref{figProp}).

\begin{figure}
%\centerline{\includegraphics[width=0.4\textwidth]{resAssBss1.eps}}
%\centerline{\includegraphics[width=0.4\textwidth]{resAssBss2.eps}}
\centerline{\includegraphics[width=0.4\textwidth]{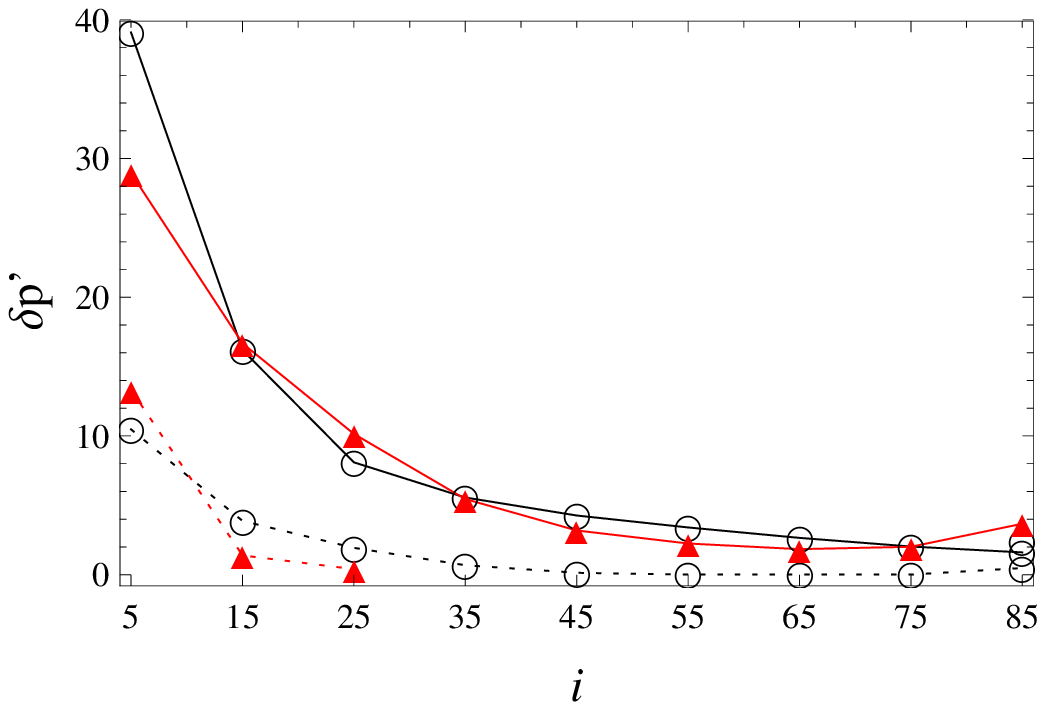}}
\centerline{\includegraphics[width=0.4\textwidth]{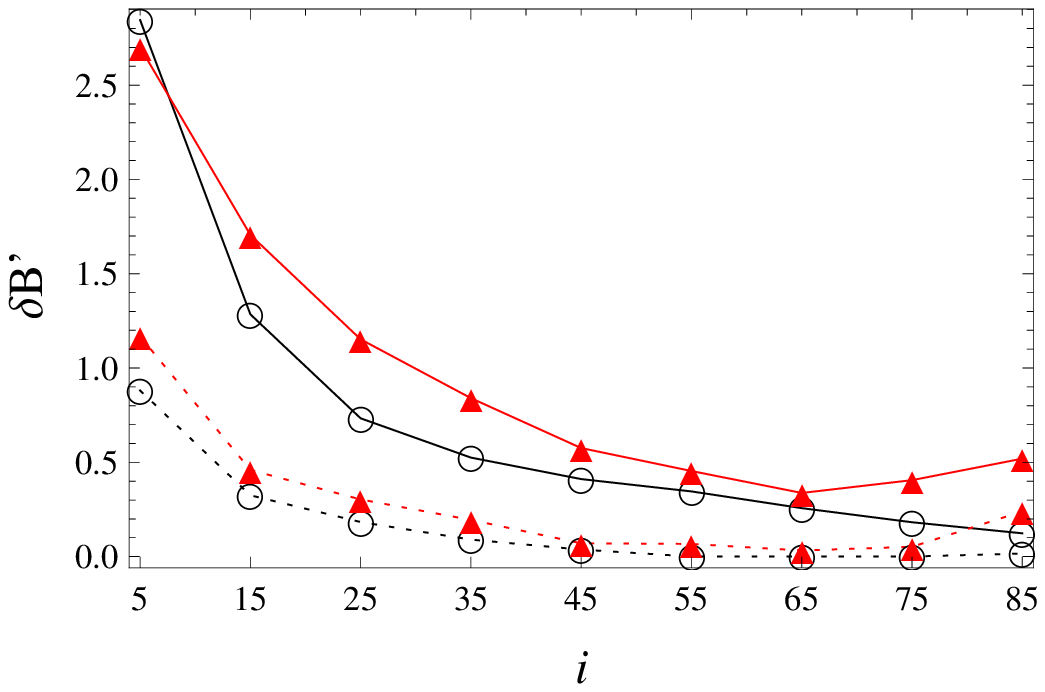}}
 \caption{Accuracy (standard deviation) of the fitted pitch angle
 (upper panel) and the magnetic field amplitude (lower panel) for the
 ASS model (open circles, black lines) and the BSS model (triangles,
 red lines) vs inclination angle $i$. Weak turbulence - dashed
 lines, strong turbulence - solid lines. Fixed parameters of the model
 are the pitch angle $p=20^{\circ}$ and the field amplitude $B=5~\mu$G.
 Each template is based on 35 $\RM$ points.}
 \label{figc5}
\end{figure}

The dependence of the fitting accuracy on the inclination angle is
shown in Fig.~\ref{figc5}. We learn that face-on galaxies always
pose more problems for recognition of the field structure. The
accuracy of the ASS or BSS recognition increases monotonically with
inclination angle due to stronger line-of-sight components of the
regular field. For inclination angles $\ge 70\degr$, the number of
effective points decreases. Furthermore, $\RM$ becomes sensitive to
fluctuations of the regular field, e.g. due to spiral arms (see
Sect.~\ref{sec:disc}), so that the reconstruction method
(Sect.~\ref{sec:deconv}) is recommended for strongly inclined
galaxies.

To illustrate the potential of identification of different models of
the regular magnetic field, using a set of templates, we present in
Table~\ref{tab:khi2} the results of recognition under a moderate
level of galactic turbulence ($\Delta \RM_{turb}=30\FRM$) for a
sample of 35 sources. The models of magnetic field structures with
fixed parameters of $p=20^{\circ}$, $\phi_0=0^\circ$, and $B=5~\mu$G
are used. The assumed level of galactic turbulence is $\Delta
\RM_{turb}=30\FRM$, the inclination of the galaxy's disk is
$45^{\circ}$, the slope of the source counts $\gamma=0.7$
(``pessimistic case''), and the ``observation parameter'' is
$Q=0.04$. The statistics for each case has been done over 220 random
distributions of 35 points within the solid angle of the galaxy and
30 different turbulent fields for each distribution (in total
220x30=6600 realizations). We adopt a significance level of 98\% for
the rejection of a model, corresponding to the $\chi^2$ probability
function Q$(\nu/2,\chi^2\nu/2)<0.02$ where $\nu= N_w - d$ is the
weighted number of degrees of freedom (Eq.~(\ref{chi2})) and $N_w
\simeq 27$ for $i=45^{\circ}$ (Fig.~\ref{figNeff}). A model is
rejected at the 98\% confidence limit if $\chi^2>1.7$ (see
Table~\ref{tab:khi2}).

In all cases the model is clearly identified as the template with
the lowest value of $\chi^2$, almost reaching the limit of 1.0 for a
perfect fit. The pitch angle is reproduced almost precisely, with an
uncertainty that decreases from $\pm 2^{\circ}$ for the ASS field to
$\pm 1^{\circ}$ for the QSS field. The accuracy of the field
amplitude is about 10\% for all field modes.

\begin{table*}
\begin{center}
\caption {The ``$\chi^2$ Championship''.} \label{tab:khi2}
\begin{tiny}
\begin{tabular}{|c|c|c|c|c|c|c|c|c|c|c|c|} \hline
template $\rightarrow$ & \multicolumn{3}{|c|}{ASS} &
\multicolumn{4}{|c|}{BSS} & \multicolumn{4}{|c|}{QSS} \\
\cline{2-12} model $\downarrow$   & $p'$ & $B'$ & $\chi^2$ & $p'$ &
$\phi_0'$ & $B'$ & $\chi^2$  & $p'$ & $\phi_0'$ & $B'$ & $\chi^2$
 \\ \hline
 \text{ASS} & \text{20$\pm $2} & \text{5.0$\pm $0.3} & \text{\bf 1.1$\pm $0.4} & \text{7$\pm $18} & \text{2$\pm $96} & \text{4.6$\pm $1.8} & \text{21$\pm $6} &
   \text{23$\pm $48} & \text{-6$\pm $41} & \text{5.2$\pm $0.8} & \text{13$\pm $3} \\
 \text{BSS} & \text{-22$\pm $47} & \text{0.6$\pm $0.7} & \text{13$\pm $3} & \text{21$\pm $2} & \text{14$\pm $32} & \text{5.1$\pm $0.4} & \text{ \bf 1.1$\pm $0.3} &
   \text{8$\pm $14} & \text{10$\pm $52} & \text{4.0$\pm $1.9} & \text{11$\pm $3} \\
 \text{QSS} & \text{-25$\pm $39} & \text{0.7$\pm $0.6} & \text{12$\pm $3} & \text{8$\pm $23} & \text{-8$\pm $95} & \text{3.5$\pm $1.3} & \text{9$\pm $3} &
   \text{20$\pm $1} & \text{-3$\pm $16} & \text{4.9$\pm $0.5} & \text{\bf 1.3$\pm $0.3} \\
 \text{ASS+BSS} & \text{19$\pm $6} & \text{2.5$\pm $0.4} & \text{3.8$\pm $1} & \text{19$\pm $16} & \text{13$\pm $107} & \text{3.2$\pm $1.1} & \text{6$\pm $2} &
   \text{24$\pm $47} & \text{2$\pm $37} & \text{2.9$\pm $0.7} & \text{6$\pm $2} \\
 \text{ASS+QSS} & \text{18$\pm $6} & \text{2.6$\pm $0.4} & \text{3.4$\pm $1} & \text{4$\pm $14} & \text{16$\pm $95} & \text{3.0$\pm $1.5} & \text{8$\pm $3} &
   \text{10$\pm $35} & \text{10$\pm $33} & \text{3.3$\pm $1.2} & \text{6$\pm $2}
\\   \hline
\end{tabular}

\begin{tabular}{|c|c|c|c|c|c|c|c|c|c|c|} \hline
template $\rightarrow$ & \multicolumn{5}{|c|}{ASS+BSS} & \multicolumn{5}{|c|}{ASS+QSS} \\
\cline{2-11}
  model $\downarrow$ & $p'$ & $q'$ & $\phi_0'$ & $B'$ & $\chi^2$  & $p'$ & $q'$ & $\phi_0'$ & $B'$ & $\chi^2$
 \\ \hline
 \text{ASS} & \text{20$\pm $2} & \text{0.9$\pm $0.1} & \text{-3$\pm $98} & \text{5.6$\pm $0.3} & \text{\bf 1.2$\pm $0.4} & \text{20$\pm $3} & \text{0.9$\pm $0.1} &
   \text{7$\pm $52} & \text{5.7$\pm $0.4} & \text{\bf 1.2$\pm $0.3} \\
 \text{BSS} & \text{21$\pm $2} & \text{0.0$\pm $0.1} & \text{9$\pm $26} & \text{5.0$\pm $0.4} & \text{\bf 1.2$\pm $0.4} & \text{8$\pm $14} & \text{0.1$\pm $0.1} & \text{2$\pm
   $50} & \text{5$\pm $2} & \text{10$\pm $3} \\
 \text{QSS} & \text{2$\pm $19} & \text{0.1$\pm $0.1} & \text{-21$\pm $95} & \text{3.7$\pm $1.2} & \text{10$\pm $4} & \text{20$\pm $1} & \text{0.0$\pm $0.1} &
   \text{0$\pm $6} & \text{5.1$\pm $0.3} & \text{\bf 1.2$\pm $0.3} \\
 \text{ASS+BSS} & \text{22$\pm $3} & \text{0.5$\pm $0.1} & \text{18$\pm $46} & \text{5.1$\pm $0.4} & \text{\bf 1.2$\pm $0.4} & \text{18$\pm $10} & \text{0.6$\pm $0.1} &
   \text{6$\pm $50} & \text{4.0$\pm $0.8} & \text{3.7$\pm $0.9} \\
 \text{ASS+QSS} & \text{15$\pm $10} & \text{0.7$\pm $0.1} & \text{-3$\pm $96} & \text{4.1$\pm $0.8} & \text{3.3$\pm $1.3} & \text{21$\pm $2} & \text{0.5$\pm $0.1} &
   \text{6$\pm $27} & \text{4.9$\pm $0.5} & \text{\bf 1.2$\pm $0.3}
\\   \hline

\end{tabular}
\end{tiny}
\end{center}
%\medskip
{Parameters of fitted pitch angles $p'$ and field amplitudes $B'$,
azimuthal phases $\phi_0'$, and $\chi^2$ values of the best-fit
template (first row). The models of field structures are given in
the first column. For the superposition templates (second part of
the Table), $q'$ is the relative amplitudes of the ASS mode, with
$q=0.5$ assumed in the model. The models with $\chi^2<1.7$ which
cannot be statistically rejected are marked in bold.}
\end{table*}

\subsection{Influence of vertical magnetic field}

We also tested models including a vertical magnetic field in the
halo, as described in Sect.~\ref{sec:rmmodels} and shown in
Fig.~\ref{fig:tilt}. We find that such a vertical magnetic field in
the halo only slightly changes the $\RM$ map even at small
inclination angles.

\begin{figure}
%\centerline{\includegraphics[width=0.4\textwidth]{resBz_p.eps}}
%\centerline{\includegraphics[width=0.4\textwidth]{resBz_B.eps}}
\centerline{\includegraphics[width=0.4\textwidth]{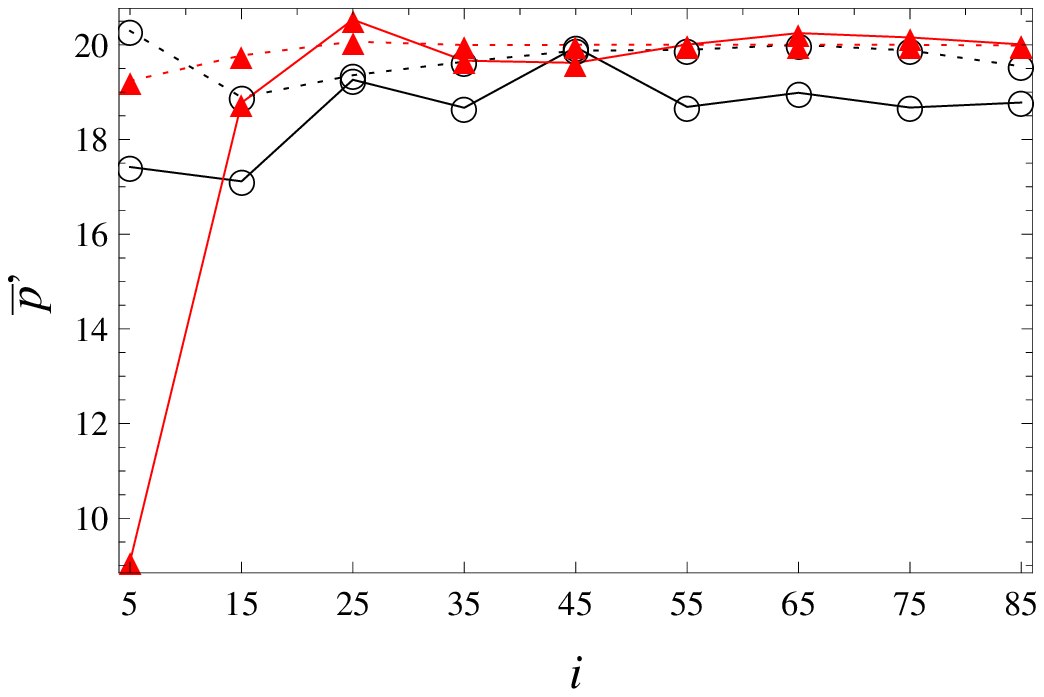}}
\centerline{\includegraphics[width=0.4\textwidth]{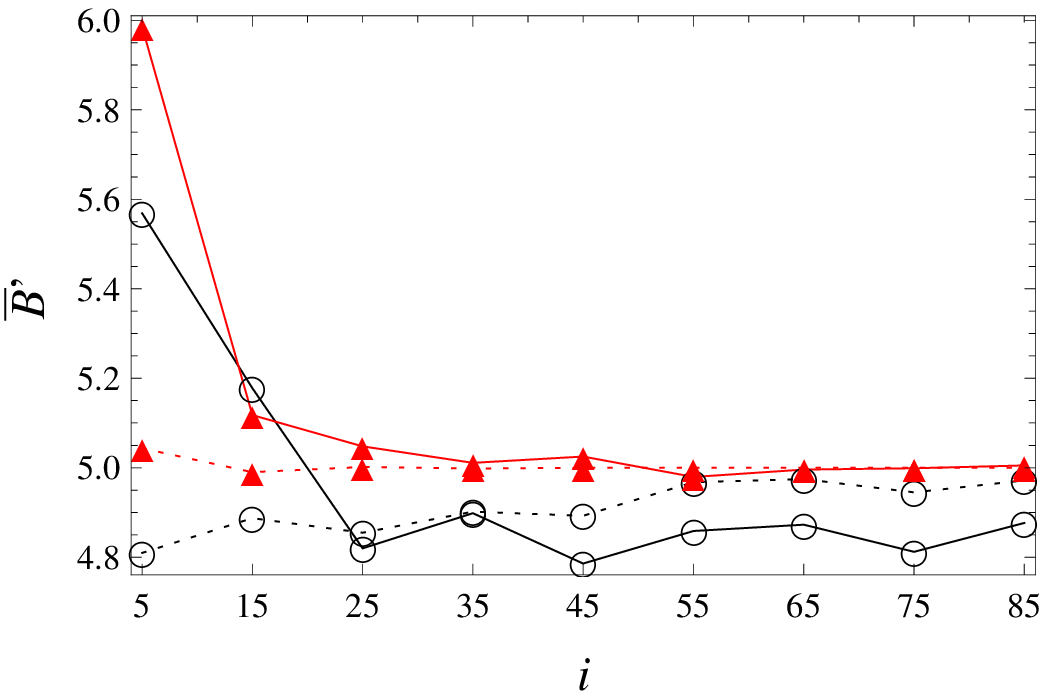}}
 \caption{Fitted pitch angle (upper panel)
 and the magnetic field amplitude (lower panel) for the ASS model
 (open circles, black lines) and the BSS model (triangles, red
 lines), both including a vertical field, recognized with
 templates without a vertical field, plotted vs inclination angle $i$.
 Weak turbulence - dashed lines, strong turbulence - solid lines.
 Fixed parameters of the model are the pitch angle $p=20^{\circ}$
 and the field amplitude $B=5~\mu$G. Each template is based on
 35 $\RM$ points.}
 \label{resBz}
\end{figure}

The influence of the vertical component of magnetic field on the
recognition procedure is studied by the following test. We calculate
the $\RM$ map for a model {\it with} a vertical magnetic field and
try to recognize it by a template having only horizontal magnetic
fields. As shown in Fig.~\ref{resBz}, the fitted pitch angles and
amplitudes of the magnetic field are in the same range as without a
vertical magnetic field for galaxies with inclination angles $i
\gtrsim 15 \degr$, but large deviations occur for smaller
inclinations. The reason becomes clear from Fig.~\ref{fig:tilt}: For
small inclinations the line-of-sight components of the vertical
field are larger than those of the field in the plane and hence
distort the results, while at larger inclinations the vertical field
becomes unimportant. We conclude that the regular magnetic field
structure of spiral galaxies (at least for inclinations of $i
> 15\degr$) can be recognized well by a magnetic field template
without a vertical field, even if the galaxy has such a vertical
magnetic field.

\subsection{Effect of position uncertainty}

Our method assumes that the position of the galaxy center known is
precisely. However, this may not be the case for distant galaxies
for which only low-resolution optical images from surveys are
available. A 2\arcsec\ uncertainty corresponds to 0.5~kpc at 50~Mpc
and 1~kpc at 100~Mpc distance.

To test the effect of position uncertainties, we shifted the
templates by 0.5~kpc and 1~kpc with respect to the models of type
ASS, BSS, and QSS. The best fits have the same parameters of the
field structures as in Table~\ref{tab:khi2}, but the $\chi^2$ values
for the ASS field increase to 1.3 and 1.8 for 0.5~kpc and 1~kpc
shifts, respectively, to 1.5 and 2.4 for the BSS field, and to 2.2
and 4.3 for the QSS field. This tells us that the disturbing effect
of a position uncertainty increases with mode number. A typical
2\arcsec\ uncertainty is uncritical for ASS-type fields even at
100~Mpc distance, while QSS-type fields can no longer be recognized
at distances greater than about 50~Mpc.

\subsection{Application of the recognition method to M~31}
\label{sec:m31}

\begin{figure}
%\centerline{\includegraphics[width=0.48\textwidth]{dataM31.eps}}
\centerline{\includegraphics[width=0.48\textwidth]{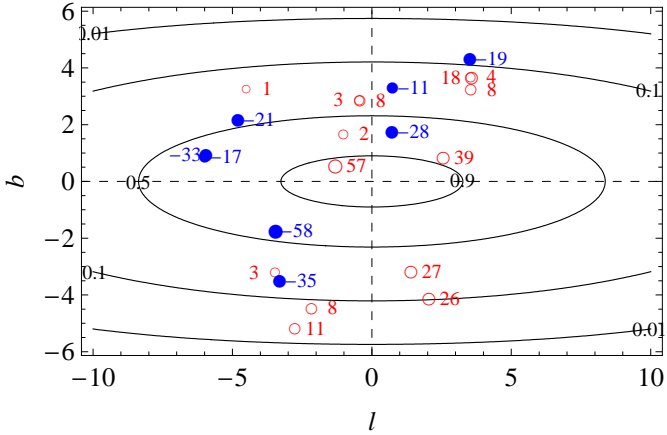}}
 \caption{$\RM$ source sample of M\,31 (from \citet{han98}).
 Numbers give $\RM$ (in $\FRM$),
 after subtraction of the mean $\RM$ from the Galactic foreground.
 Isolines show the weight function used for $\chi^2$. Coordinates are
 given in kpc.}
 \label{fig:M31}
\end{figure}

\begin{table*}
\begin{center}
\caption {The ``$\chi^2$ Championship'' for M\,31.} \label{tab:m31m}
\begin{tiny}
\begin{tabular}{|c|c|c|c|c|c|c|c|c|c|c|c|} \hline
template $\rightarrow$ & \multicolumn{3}{|c|}{ASS} &
\multicolumn{4}{|c|}{BSS} & \multicolumn{4}{|c|}{QSS} \\
\cline{2-12}
  model $\downarrow$  & $p'$ & $B'$ & $\chi^2$ & $p'$ & $\phi_0'$ & $B'$ & $\chi^2$  & $p'$ & $\phi_0'$ & $B'$ & $\chi^2$
 \\ \hline
  \text{ASS} & \text{20$\pm $1} & \text{5.0$\pm $0.2} & \text{\bf 1.3$\pm $0.6} & \text{14$\pm $1} & \text{128$\pm $4} & \text{8.8$\pm $0.5} & \text{84$\pm $7} &
   \text{62$\pm $6} & \text{-5$\pm $3} & \text{4.2$\pm $0.2} & \text{71$\pm $10} \\
 \text{BSS} & \text{0$\pm $58} & \text{0.0$\pm $0.1} & \text{76$\pm $8} & \text{20$\pm $1} & \text{0$\pm $3} & \text{5.0$\pm $0.2} & \text{\bf 1.4$\pm $0.6} &
   \text{15$\pm $2} & \text{11$\pm $35} & \text{9.1$\pm $0.9} & \text{34$\pm $5} \\
 \text{QSS} & \text{-2$\pm $31} & \text{0.4$\pm $0.2} & \text{14$\pm $3} & \text{19$\pm $11} & \text{-122$\pm $77} & \text{3.7$\pm $3.5} & \text{9$\pm $2} &
   \text{20$\pm $1} & \text{0$\pm $3} & \text{5.1$\pm $0.5} & \text{\bf 1.5$\pm $0.5} \\
 \text{ASS+BSS} & \text{30$\pm $5} & \text{1.5$\pm $0.2} & \text{13$\pm $2} & \text{14$\pm $1} & \text{165$\pm $4} & \text{4.8$\pm $0.5} & \text{20$\pm $3} &
   \text{58$\pm $8} & \text{-10$\pm $31} & \text{1.9$\pm $0.4} & \text{13$\pm $2} \\
 \text{ASS+QSS} & \text{20$\pm $3} & \text{2.6$\pm $0.2} & \text{4.8$\pm $1.4} & \text{14$\pm $1} & \text{123$\pm $17} & \text{4.7$\pm $0.4} & \text{25$\pm $4} &
   \text{58$\pm $19} & \text{-6$\pm $10} & \text{2.5$\pm $1.4} & \text{24$\pm $4}
\\   \hline
\end{tabular}

\begin{tabular}{|c|c|c|c|c|c|c|c|c|c|c|} \hline
template $\rightarrow$ & \multicolumn{5}{|c|}{ASS+BSS} & \multicolumn{5}{|c|}{ASS+QSS} \\
\cline{2-11}
  model $\downarrow$ & $p'$ & $q'$ & $\phi_0'$ & $B'$ & $\chi^2$  & $p'$ & $q'$ & $\phi_0'$ & $B'$ & $\chi^2$
   \\ \hline
 \text{ASS} & \text{20$\pm $1} & \text{0.9$\pm $0.1} & \text{15$\pm $112} & \text{5.6$\pm $0.3} & \text{\bf 1.7$\pm $0.8} & \text{20$\pm $1} & \text{0.9$\pm $0.1} &
   \text{5$\pm $58} & \text{5.6$\pm $0.3} & \text{\bf 1.8$\pm $0.7} \\
 \text{BSS} & \text{20$\pm $1} & \text{0.0$\pm $0.1} & \text{0$\pm $5} & \text{5.0$\pm $0.3} & \text{\bf 1.9$\pm $0.9} & \text{15$\pm $2} & \text{0$\pm $0.1} & \text{7$\pm $39}
   & \text{9$\pm $1} & \text{43$\pm $5} \\
 \text{QSS} & \text{14$\pm $13} & \text{0.0$\pm $0.1} & \text{-121$\pm $82} & \text{5.2$\pm $3.9} & \text{11.4$\pm $2.5} & \text{20$\pm $1} & \text{0.0$\pm $0.1} &
   \text{0$\pm $1} & \text{5.1$\pm $0.6} & \text{\bf 1.5$\pm $0.5} \\
 \text{ASS+BSS} & \text{20$\pm $2} & \text{0.5$\pm $0.1} & \text{5$\pm $27} & \text{5.0$\pm $0.3} & \text{\bf 1.5$\pm $0.7} & \text{25$\pm $5} & \text{0.4$\pm $0.1} &
   \text{0$\pm $12} & \text{4.4$\pm $1.5} & \text{8$\pm $2} \\
 \text{ASS+QSS} & \text{20$\pm $4} & \text{0.7$\pm $0.1} & \text{83$\pm $125} & \text{3.7$\pm $1.2} & \text{4.2$\pm $1.7} & \text{20$\pm $1} & \text{0.5$\pm $0.1} &
   \text{2$\pm $13} & \text{5.1$\pm $0.6} & \text{\bf 1.5$\pm $0.5}
\\   \hline
\end{tabular}
\end{tiny}
\end{center}
%\medskip
{Parameters of fitted pitch angles $p'$ and field amplitudes $B'$,
azimuthal phases $\phi_0'$, and $\chi^2$ of the best fit for
different model fields (first column), recognized by a template
(first row) based on the existing sample of 22 sources in the M~31
area. Models with $\chi^2>2.7$ (marked in bold) are rejected at the
98\% significance level.}
\end{table*}

\cite{han98} measured 22 $\RM$ values from polarized background
sources within a solid angle of about 1~deg$^2$ in the inner part of
M~31 (Fig.~\ref{fig:M31}). Their VLA observations at $\lambda$=22~cm
and at $\lambda$=18~cm ($\Delta \lambda / \lambda \approx 0.015$)
reached a sensitivity limit of 0.3~mJy/beam area. The source number
agrees well with the source counts from Fig.~\ref{fig:ncp}.
\cite{han98} claimed that their $\RM$ values are consistent with an
ASS field in M~31, but did not include a statistical test.

Observations of the diffuse polarized emission and its $\RM$ from
M~31 has been described by an ASS field with $8^{\circ}-19^{\circ}$
pitch angle, $4~\mu$G amplitude (assuming $\n=0.03\cmcube$ and 1~kpc
pathlength), with a weaker QSS field superimposed in the inner part
of the galaxy \citep{fletcher04}, while earlier results indicated a
superimposed weaker BSS field \citep{sofue87}.

Firstly, we estimate how representative the sample of \cite{han98}
is. We use the same models with fixed parameters and assumptions as
described in Sect.~\ref{sec:results} and repeat the same test as
described above (see Table~\ref{tab:khi2}) for different model
fields, but using for the recognition only the coordinates of the 22
sources (shown in Fig.~\ref{fig:M31}) and the observed instrumental
errors $\Delta\RM_{noise}$ for each source, but we used the
$\RM$ values obtained from the field models. The results in
Table~\ref{tab:m31m} show that the sample of 22 sources is large
enough to clearly recognize the field structure with simple symmetry
(if it exists). Models with $\chi^2>2.7$ are rejected at the 98\%
significance level, which corresponds to a weighted number of
degrees of freedom of $\nu=N_w-d \simeq 5$, where $N_w \simeq 8$
(for the inclination of M~31 of $i=75^{\circ}$) and the number of
free parameters is $d=2$, 3, or 4.

Secondly, we use the observed data to apply our recognition model.
The analysis was performed in a first run for all 22 points. In a
second run we exclude the two sources with the highest $\RM$
deviations from the template model field (those with
+57~rad~m$^{-2}$ and -17~rad~m$^{-2}$), following the culling
algorithm as described in Sect.~\ref{sec:models} in order to reduce
the influence of strong internal $\RM$ of the background sources. As
shown in Table~\ref{tab:m31r} none of the two runs give a clear
result. Using all 22 sources, all templates give a similar value of
$\chi^2$. Taking only 20 sources (middle part of
Table~\ref{tab:m31r}), a small preference for the ASS model is
indicated. However, all models are rejected at the 98\% significance
level ($\chi^2>2.7$). As the simple field structures could be
clearly identified in our test using the same set of sources
(Table~\ref{tab:m31m}), the results of Table~\ref{tab:m31r} seem to
indicate that the structure of the regular magnetic field of M~31 is
more complicated than suggested by our templates. However, the
analysis of the diffuse emission by \citep{fletcher04} is
statistically reliable and does not require higher modes with
significant amplitudes. We conclude that the measured $\RM$ values
towards the background sources contain a significant, intrinsic
contribution or a fluctuating contribution of the Milky Way
foreground, which can be subtracted only if a larger number of
background sources becomes available.

Thirdly, to account for the $\RM$ intrinsic to the sources or from
the Galactic foreground, we increase all $\RM$ errors by an
arbitrary factor of three. The $\chi^2$ values of the best fit
decrease and the ASS model becomes slightly preferred statistically
(see bottom part of Table~\ref{tab:m31r}), in agreement with the
results from diffuse $\RM$.

\begin{center}
\begin{table}
\caption {The field recognition for M\,31.} \label{tab:m31r}
\begin{tiny}
\begin{tabular}{|c|c|c|c|c|c|c|} \hline
  N  & temp. & $p'$ & $q'$ & $\phi_0'$ & $B'$ & $\chi^2$  \\ \hline
  & \text{ASS} & \text{-8$\pm $31} & - & - & \text{0.4$\pm $0.2} & \text{5.3$\pm $1.2} \\
  & \text{BSS} & \text{12$\pm $16} & - & \text{46$\pm $121} & \text{2.2$\pm $2.6} & \text{5.0$\pm $1.1} \\
 22 & \text{QSS} & \text{3$\pm $45} & - & \text{-10$\pm $53} & \text{1.7$\pm $2.1} & \text{5.0$\pm $1.2} \\
  & \text{A+B} & \text{6$\pm $16} & \text{0.1$\pm $0.1} & \text{-51$\pm $115} & \text{2.7$\pm $1.9} & \text{5.7$\pm $1.4} \\
  & \text{A+Q} & \text{44$\pm $34} & \text{0.3$\pm $0.1} & \text{-39$\pm $38} & \text{2.6$\pm $1.7} & \text{5.1$\pm $1.2} \\
 \hline
  & \text{ASS} & \text{-16$\pm $22} & - & - & \text{0.4$\pm $0.2} & \text{4.6$\pm $1.2} \\
  & \text{BSS} & \text{1$\pm $21} & - & \text{85$\pm $92} & \text{3.7$\pm $3.6} & \text{5.8$\pm $1.3} \\
 20 & \text{QSS} & \text{9$\pm $36} & - & \text{-12$\pm $52} & \text{2.4$\pm $2.7} & \text{5.1$\pm $1.2} \\
  & \text{A+B} & \text{-17$\pm $17} & \text{0.4$\pm $0.2} & \text{34$\pm $89} & \text{2.$\pm $1.5} & \text{6.3$\pm $1.7} \\
  & \text{A+Q} & \text{8$\pm $32} & \text{0.2$\pm $0.2} & \text{-28$\pm $50} & \text{2.8$\pm $2.1} & \text{6.4$\pm $1.6} \\
  \hline
  & \text{ASS} & \text{-3$\pm $17} & - & - & \text{0.5$\pm $0.2} & \text{\bf 2.8$\pm $0.9} \\
  & \text{BSS} & \text{8$\pm $15} & - & \text{17$\pm $120} & \text{2.8$\pm $2.4} & \text{3.3$\pm $1.2} \\
 \text{$\tilde{20}$} & \text{QSS} & \text{28$\pm $39} & - & \text{-12$\pm $42} & \text{2.6$\pm $3.1} & \text{3.3$\pm $1.1} \\
  & \text{A+B} & \text{3$\pm $19} & \text{0.3$\pm $0.2} & \text{9$\pm $114} & \text{3.1$\pm $2.3} & \text{3.8$\pm $1.4} \\
  & \text{A+Q} & \text{19$\pm $34} & \text{0.3$\pm $0.1} & \text{-21$\pm $49} & \text{3.6$\pm $2.1} & \text{3.7$\pm $1.2}
 \\ \hline
\end{tabular}
\end{tiny}

\medskip
{Parameters of fitted pitch angles $p'$, field amplitudes $B'$,
azimuthal phases $\phi_0'$, amplitude ratios $q'$ (in case of a
superposition), and $\chi^2$ of the best fit, using 22 and 20 $\RM$
values observed in the M~31 regime by \cite{han98}. For 22 and 20
values (8 and 6.5 effective points), models with $\chi^2>2.7$ and
$\chi^2>3.1$, respectively, are rejected at the 98\% significance
level.}

\end{table}
\end{center}

%----------------------------------------------------------------------

\section{Magnetic field reconstruction from $\RM$ grids of inclined galaxies}
 \label{sec:deconv}

In the previous section the possibility of recognizing the
large-scale structure of the regular magnetic field using a small
number of sources has been discussed. In this one we switch to the
case of a large number of sources, which allows us to reconstruct
the structure of the galactic field, without precognition of a given
set of simple models. The method discussed below is especially
powerful for resolving the detailed structure of the magnetic field
of a strongly inclined galaxy, provided an $\RM$ grid with strong
enough resolution is available.

Observations reveal a grid of $\RM(l,b)\pm\Delta\RM(l,b)$ where the
standard deviation $\Delta\RM(l,b)$ characterizes the observational
noise. Note again that here, in contrast to the recognition method
discussed in the previous section, we do not separate $\RM(l,b)$
into its {\it regular} and {\it turbulent} parts, but instead merely
reconstruct the existing field structure, trying to suppress the
influence of random signals (instrumental or generated in the
galaxy). The Faraday rotation measure is proportional to the product
$\left( \Brpa \, n_e \right)$. Ignoring the problem of the $\Brpa$
and $n_e$ separation, we consider the problem of reconstructing the
magnetic field $\left( \Brpa \, n_e \right)$ from the observed $\RM$
grid. We suppose that $\left(  \Brpa \, n_e \right)$ can be written
in the form:

\begin{equation}
\left( \Brpa \, n_e \right) = f(x,y) \, \, \eta(z). \label{fh}
\end{equation}
This allows us to describe the decrease in $|{\rm RM}|$ in the
$z$-direction by one term $\eta(z)$ and to state the problem as the
reconstruction of the 2-D scalar function $f(x,y)$ (see
Fig.~\ref{fig:coord} for the coordinate system).

The aim is to reconstruct the field $f(x,y)$ from the $\RM$ grid.
The rotation measure grid, even in the case of
%ideal resolution and
noiseless measurements, is a convolution of the required field
$f(x,y)$ with some function $g(y)$, defined by the vertical
structure of the galactic field. For the case $B_z=0$, the function
$g(y)$ is defined only by the function $\eta(z)$. Hence,

\begin{eqnarray}\label{RM2}
 && \RM(l,b) = \nonumber \\ &&  = C \int_{-\infty}^{\infty}
\int_{-\infty}^{\infty} f(x,y) \, \eta(z) \, \delta
\left(z-\frac{y\cos i -b}{\sin i}\right) dy dz = \nonumber
\\
&& = C \int_{-\infty}^{\infty} f(x,y) \, g(b/\sin i -y) \, dy,
\end{eqnarray}
where $$g(y)= \int_{-\infty}^{\infty}\eta\left(y\cos i\right) \,
dz,$$ where $i$ is the inclination angle and $C$ a constant.

The standard solution for a problem like Eq.~(\ref{RM2}) is based on
the Fourier deconvolution. Actually,
\begin{equation}
\label{decF}
 \hat f(x, k)  = \frac{\hat{\RM}(l, k)}{\hat g(k)}
\end{equation}
where $k$ is the wave number. The hat denotes the Fourier transform
of the corresponding function. The Fourier deconvolution gives the
exact solution of Eq.~(\ref{RM2}), but requires some regularization
procedure \citep{heinz-martin96} to avoid the amplification of high
spatial frequency noise, especially if the spectrum of the kernel
function $g$ decreases rapidly with $k$. The often-suggested
Gaussian form of the galactic profile $\eta(z)$ leads to a kernel
that is ``bad'' in this sense. We do not discuss the regularization
problem here in detail. The simplest algorithms use some kind of
high-frequency cut-off in the Fourier space, while more
sophisticated ones introduce subtle filtration, based for example on
wavelet algorithms \citep{decon04}.

The main problems of using Eq.~(\ref{decF}) arise from the limited
resolution of the $\RM$ grid (small number of points), noise, and
unknown kernel function $g$ (namely, one does not know the vertical
distribution function $\eta(z)$).

\begin{figure}
\begin{center}
 \includegraphics[width=80mm]{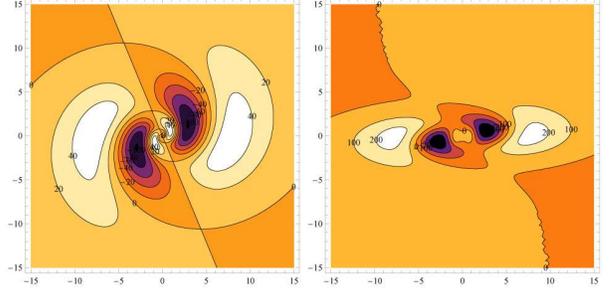}
  \caption{The function $f(x,y)$ (left) and the $\RM$ grid (right) for a
  BSS galaxy model (pitch angle
  $p=20^{\circ}$) inclined at $i=70^{\circ}$ .}
  \label{fig:rm_bss_70}
\end{center}
\end{figure}

\begin{figure}
\begin{center}
 \includegraphics[width=70mm]{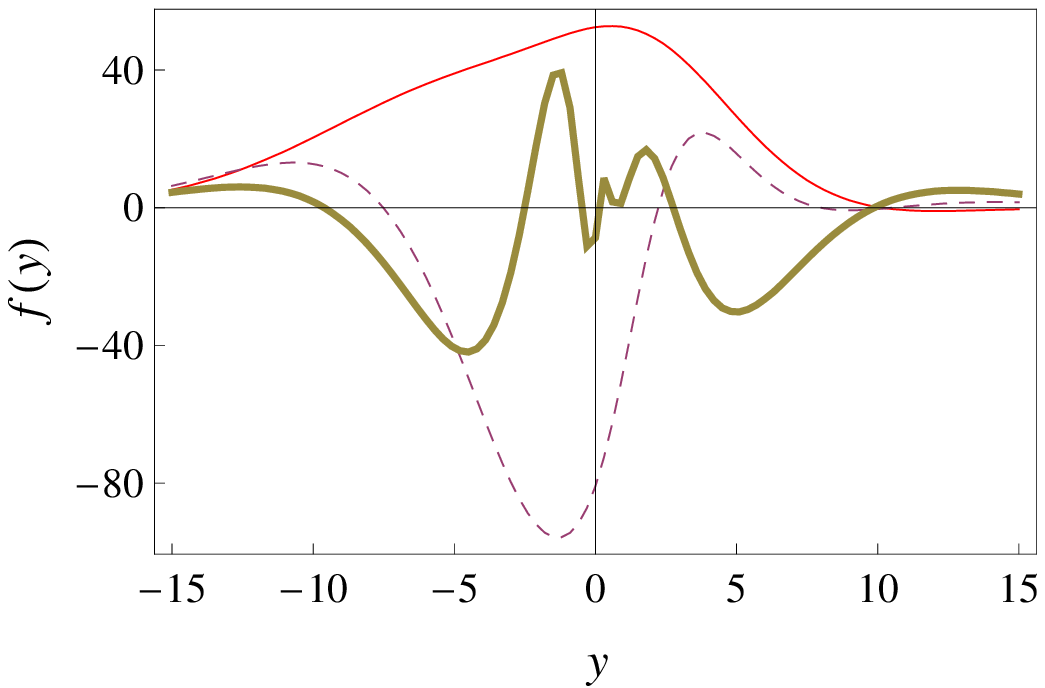}
 \includegraphics[width=70mm]{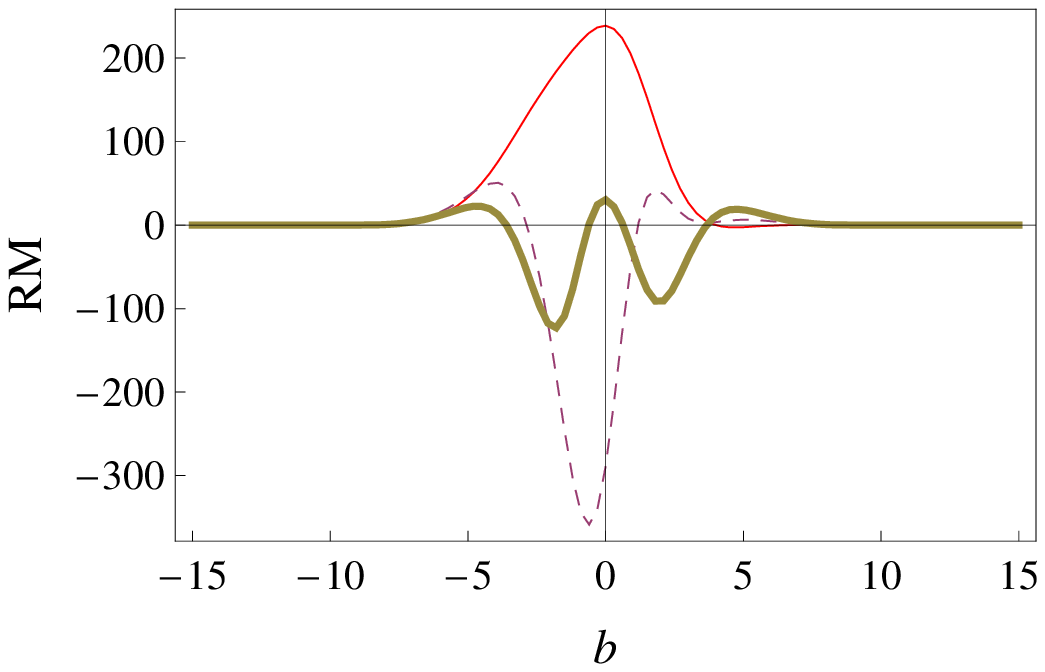}
  \caption{The profiles $f(y)$ along or parallel to the minor axis of the
  projected galaxy disk in the sky plane and the corresponding profiles of
  $\RM(b)$ for a BSS-type galaxy, inclined at $i=70^{\circ}$: $x=0$
  (thick curve), $x=3$~kpc (dashed), and $x=7.5$~kpc (thin).}
  \label{fig:rm_bss}
\end{center}
\end{figure}

As an example let us consider a strongly inclined galaxy with a pure
BSS-type magnetic field. The required function $f(x,y)$ for this
case and the simulated $\RM$ grid for the inclination angle
$i=70^{\circ}$ are shown in Fig.~\ref{fig:rm_bss_70}.
The problem can be illustrated by reconstructing the one-dimensional
profiles $f(y)$ for given cross-sections $x=const$. Figure
\ref{fig:rm_bss} shows the profiles of $f(y)$ for three different
values of $x$ and the corresponding three profiles of $\RM(b)$ at
the inclination $i=70^{\circ}$.
To use the deconvolution (\ref{decF}) one needs the function
$\eta(z)$. Even if a Gaussian shape is adopted, one free parameter
(the half-thickness of the ionized gas disk $h$) remains. We denote
by $h'$ the supposed value of $h$ used for the deconvolution.

\begin{figure}
\begin{center}
 \includegraphics[width=70mm]{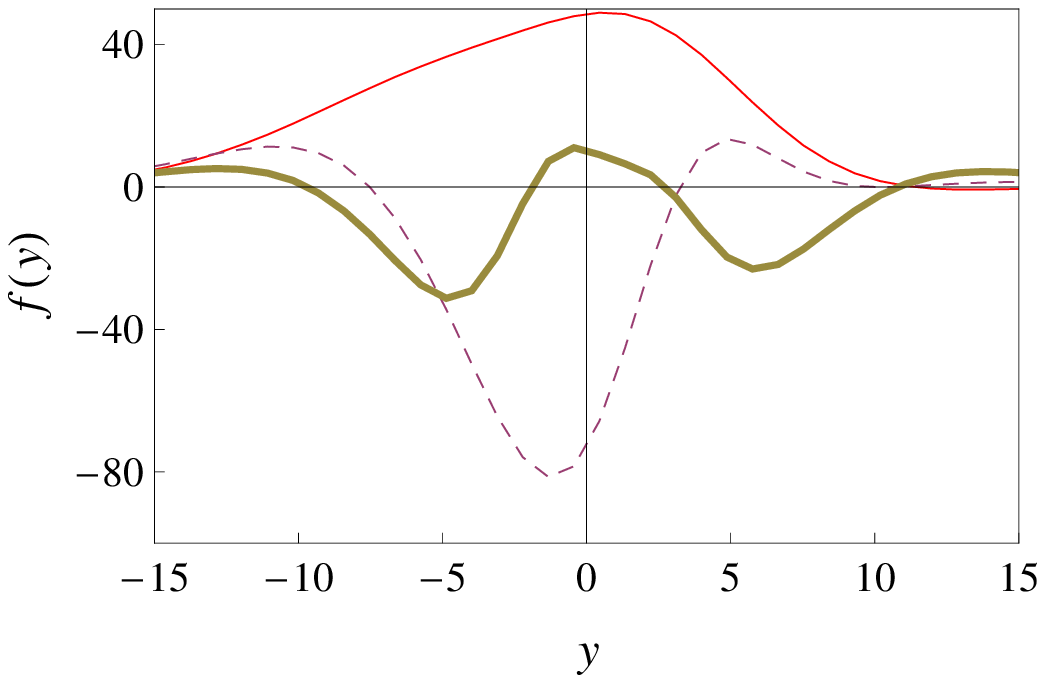}
 \includegraphics[width=70mm]{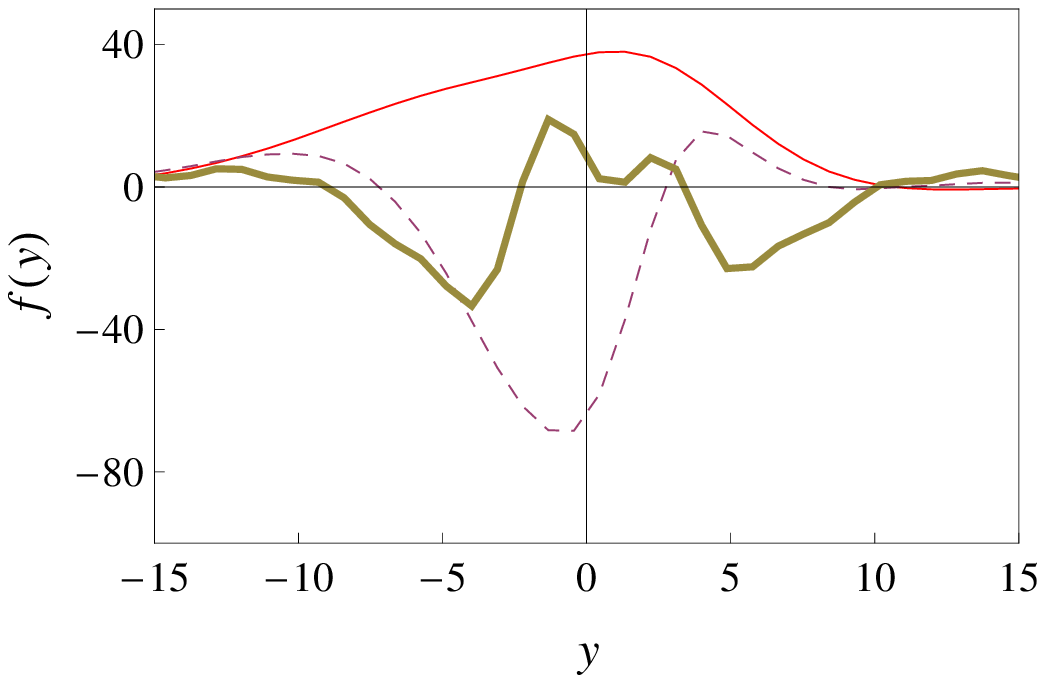}
 \includegraphics[width=70mm]{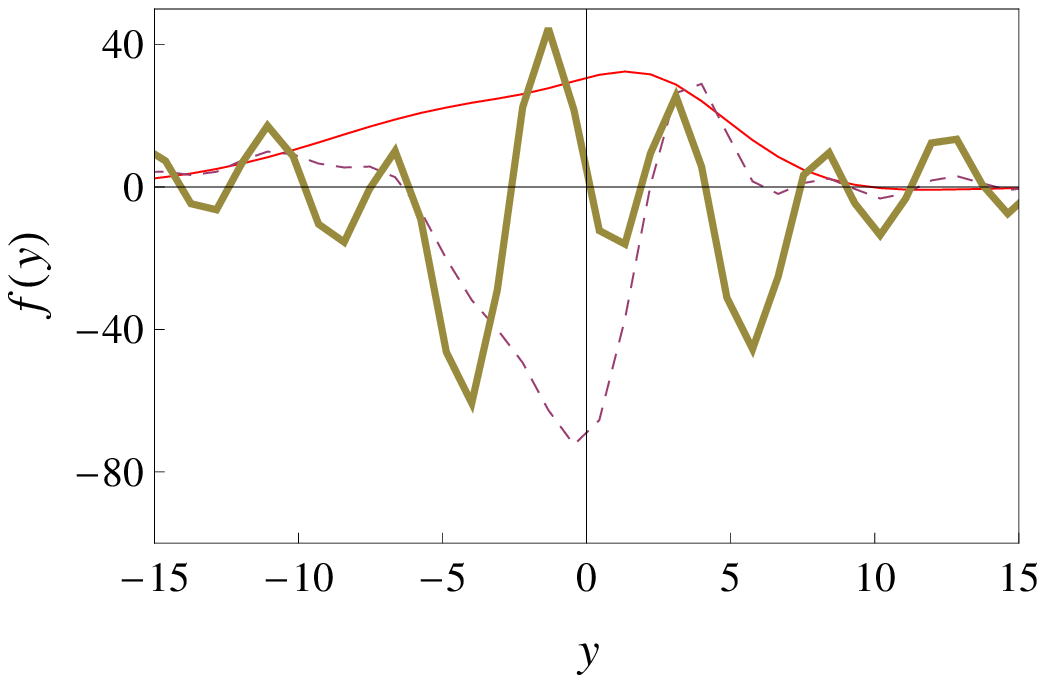}
 \caption{The profiles $f(y)$ reconstructed using three different
   values of the supposed galactic thickness: $h'=0.75$~kpc (top),
   $h'=1$~kpc (middle), $h'=1.25$~kpc (bottom). The correct value is
   $h=1$~kpc. The profiles are given for $x=0$ (thick curve),
   $x=3$~kpc (dashed), and $x=7.5$~kpc (thin).
   }
 \label{fig:rm_bss_r}
\end{center}
\end{figure}

Figure \ref{fig:rm_bss_r} shows the profiles $f(y)$ reconstructed
from $\RM(b)$ by deconvolution with different assumed values for the
half thickness $h'$. The simulation is done for a set of points
corresponding to 20 sources across the galaxy (which seems to be
quite realistic in the frame of the SKA project). This figure
illustrates that the result of reconstruction strongly depends on
the choice of $h'$. An underestimated value of the thickness of the
ionized disk leads to insufficient reinforcement of small-scale
structures -- the deconvolution merely stretches the $\RM$
projection. In contrast, an overestimation of $h'$ implies an
overrated smoothing and results in a disproportional amplification
of high spatial frequencies (small scales). According to the last
panel of Fig.~\ref{fig:rm_bss_r}, the oscillations of $f(y)$
essentially exceed the real solid angle of the galaxy. The correct
value $h'=h$ allows restoration of two positive maxima in the
central part of the minor axis profile $x=0$, which were completely
hidden in the $\RM$ projection. This example tells us that a slight
underestimate of $h'$ is more acceptable than an overestimate.
Figure \ref{fig:rm_bss_r} also shows that the central profile at
$x=0$ is the most informative one for the reconstruction, while the
outer profiles are more similar.

%==================================================================

\section{Discussion}
\label{sec:disc}

Up to now the analysis of the magnetic field structure in external
galaxies from $\RM$ observations is still in a rudimentary state.
There are only two nearby galaxies behind which a limited sample of
polarized background sources were observed (M~31 and LMC). The
impressive projects of large radio telescope arrays (LOFAR and SKA)
promote the investigations of future possibilities. We considered
two kinds of analyzes that may become possible: the detailed {\em
reconstruction} of the structure of the regular magnetic field in a
number of nearby galaxies from a dense grid of $\RM$ values from
background sources in the galaxy field, and the {\em recognition} of
the main symmetry of the large-scale regular magnetic field from the
$\RM$ of a number of background sources. We studied this problem
based on current estimates of the increasing number density of
polarized sources with improving telescope sensitivity. We took an
optimistic and a pessimistic extrapolation of the existing
cumulative number counts into account.

In the first part of the paper, we studied the possibility of
recognizing the large-scale regular component of the magnetic field
with a simple symmetry using a straightforward testing by templates,
which correspond to axisymmetric, bisymmetric, and quadrisymmetric
modes, or their superpositions, and applying the $\chi^2$ criterium
for evaluating the reliability of the different modes. We showed
that, if a symmetric part indeed exists in the analyzed field (even
mixed with a relatively strong turbulent field), dozens of sources
provide a good chance of a reliable recognition.  These templates
can be even successfully applied if the disk field is accompanied by
an X-shaped vertical field as found in nearby galaxies observed
almost edge-on. The result strongly depends on the quality of the
observations, the flux density of the sources, and their number
density distribution.

Our results show that regular magnetic fields of simple azimuthal
symmetry (modes), like axisymmetric (ASS), bisymmetric (BSS), and
quadrisymmetric spiral fields (QSS), and superpositions of these,
can be recognized with $\RM$ values from a number of background
sources depending on the inclination of the galaxy's disk, the level
of $\RM$ turbulence in the galaxy, and the slope of the number
density distribution of the polarized sources. The fitted parameters
are the spiral pitch angle, the amplitude of the regular field
strength, and the phase (for non-axisymmetric fields). The typical
accuracies achieved for the fitted parameters are $\pm(
1^{\circ}-4^{\circ})$ for the pitch angle and $\pm10\%$ for the field
amplitude. Under the favorite conditions, $10-30$ sources are
already reliable enough to recognize the field structure. However,
the unknown {\it internal} $\RM$ of the background sources
themselves may require a larger number of measured sources in
order to statistically average out the contribution of internal RM,
or to be able to apply a culling algorithm, as mentioned in
Sect.~\ref{sec:models}.

The regular field of a higher order of symmetry can be easier
recognized; i.e. it requires fewer points and shorter observation
time for the same accuracy of the fitted pitch angle definition and
is less affected by the turbulent component of $\RM$. The BSS and
QSS fields seem to have better chances of being recognized in
slightly inclined galaxies. However, the smaller required $N^*$ and
the larger number of parameters for the higher mode decreases the
reliability of the recognition.

The dependence on turbulence becomes dramatic for weakly inclined
(almost face-on) galaxies -- a reliable fitting requires a huge
number of sources. This problem may occur especially for small,
slowly rotating galaxies with a high rate of star formation, where
the large-scale regular field is expected to be weaker than the
turbulent field \citep{beck96}. Interactions or ram pressure may
also enhance the turbulent field and mask any large-scale pattern in
the $\RM$ distribution.

In our models we did not consider deviations of the regular field
structure from simple azimuthal modes, e.g. due to spiral arms or
additional field reversals. These would lead to further $\RM$
fluctuations that may become large for strongly inclined galaxies
and hence decrease the accuracy of the fits. Strong $\RM$
fluctuations have been observed near the plane of the Milky Way
\citep{brown07}.

For our test case M~31, the best case of a dominating field mode
(ASS) so far, the available $\RM$ data from polarized background
sources confirm the results from the analysis of $\RM$ data from the
diffuse emission of M~31 (Sect.~\ref{sec:m31}) if the $\RM$ errors
are increased to account for the $\RM$ contributions intrinsic to
the sources or from the foreground in the Milky Way. A larger number
of sources is needed for proper subtraction of these effects.
However, it is possible that higher field modes exist and will
prevent a clear result even in case of a greater source number.

The detection of the ASS mode in M~31 from $\RM$ of background
sources also indicates that this mode is symmetric with respect to
the plane (mode S0) because $\RM$ in a strongly inclined galaxy
hosting an A0 mode should reverse above and below the plane. The
$\RM$ amplitude of polarized sources behind a S0-type field should
be twice larger than that from the $\RM$ map of the diffuse
polarized emission from the galaxy itself, which has to be confirmed
with more sensitive data.

For many galaxies no strongly dominating field mode can be expected.
A high $\chi^2$ value, which does not decrease with increasing
source number, would indicate that the field structure is more
complicated than a superposition of a few simple modes or that the
turbulent field is stronger than the regular field. In this case,
our {\em reconstruction} method should be applied, because it does
not need a ``precognition'' template but needs a higher density of
$\RM$ sources and hence deeper observations.

Observations provide the 2-D projection of the product
$(B_{||}~n_e)$. Interpretation of this projection can by impeded by
a vertical magnetic field $B_z$, but the contribution of this field
is not crucial in the case of symmetrically tilted fields. If the
inclination angle increases, the $\RM$ grid is entirely dominated by
the horizontal field, but another problem arises: the smoothing of
detailed structures (of the size of a spiral arm) due to their
superposition in a highly inclined projection. In other words, the
function $f(x,y)$ describing the field structure in the galaxy's
plane is convolved by a smoothing function, defined by the vertical
profile of the galaxy. Provided a universal shape of the profile
exists, one can apply the deconvolution technique to restore the
structure of the magnetic field in the galaxy's plane. In principle,
this method works for any field structure.

The reconstruction method is superior for strongly inclined galaxies
(about $70^{\circ}$ and more) and is successful if a large sample of
sources is available. At least about $20$ sources are required for
one cut along the {\it minor} galactic axis (or $\sim1200$ sources
within the solid angle covered by the galaxy) to reconstruct the
spiral arms in a spiral galaxy. The number of points increases if
the scale of reconstructed details decreases, following the
well-known Kotelnikov--Nyquist relation.

Note that the reconstruction procedure can be applied to the $\RM$
obtained from the diffusive polarized emission of the galaxy itself,
but this would require the deconvolution of a more complicated
integral equation than Eq.~(\ref{RM2}). On the other hand, the
recognition method also works for a continuous $\RM$ map obtained
from the diffuse polarized emission.

Finally, we wish to point out that little is known about the
statistical properties of the $\RM$ contributions intrinsic to the
background sources and of the contribution of small-scale $\RM$
inhomogeneities in the foreground of the Milky Way. Large, nearby
galaxies will suffer most from fluctuations in the foreground.
Another uncertainty are depolarization effects within the background
sources, or in the foreground of the galaxy or the Milky Way, if the
angular extent of a source is greater than the angular turbulence
scale in the foreground medium. Depolarization of background sources
in the LMC were detected by \citet{gaensler05}. The latter
depolarization effect increases with distance because larger
turbulence scales can contribute. Both depolarization effects,
internal and external, depend on angular resolution and on
observation frequency. No statistical data are available yet.
Further investigations in this direction may modify the number of
sources required for a reliable recognition.

\section{Application to LOFAR and SKA}
\label{sec:ska}

The {\em reconstruction} of magnetic field structures of strongly
inclined spiral galaxies is possible for a sample of $\geq 1200$
$\RM$ sources. This would require a sensitivity of the SKA at
1.4~GHz of $\approx 0.5-5~\mu$Jy (or integration time less than one
hour) for galaxies at distances of about one Mpc. The field
structures of galaxies at about 10~Mpc distance can be reconstructed
with tens to a hundred hours of integration time.

The results for the {\em recognition} method presented in this paper
are very promising for future observations with the SKA of
background sources behind galaxies that are too faint to be observed
directly by their extended polarized emission. Simple field
structures can be recognized with the SKA at 1.4~GHz (21~cm) with an
``observation parameter'' $Q\approx 1$. This needs a $\sigma_p
\approx 0.2~\mu$Jy for galaxies at 30~Mpc distance and $\sigma_p
\approx 0.015~\mu$Jy for galaxies at 100~Mpc distance. These
sensitivities can be achieved within 15~min and 100~h observation
time, respectively, assuming that polarization calibration of the
SKA is possible down to such low flux-density levels.

Let us estimate the number of spiral galaxies for which one can
reconstruct or recognize the magnetic fields. \citet{marinoni99}
estimate the mean number of galaxies ($\simeq 0.032$~Mpc$^{-3}$; the
multi-attractor model was used to predict the distances to the
galaxies) for a magnitude-limited all-sky optical sample of nearby
$\simeq 5300$ galaxies with recession velocities of
$cz<5500$~km~s$^{-1}$. The mean density of 2240 Sbc--Sd spirals is
about 0.013~Mpc$^{-3}$, which counts for about 60 spirals within a
distance of 10~Mpc and $\approx 60000$ spirals within 100~Mpc.

The recognition method raises requirements for the dynamic range of
the telescope for polarization measurements (polarization purity),
which can be directly taken from Fig.~\ref{fig:ncp}. The detection
of about 100 polarized sources in the solid angle of a galaxy at
about 1~Mpc distance needs a dynamic range of about 20~dB, while at
10~Mpc distance 30~dB may already be needed for a pessimistic value
of $\gamma$, and even 40~dB at 100~Mpc distance. Hence, reliable
estimates of $\gamma$ are essential to compute the required
polarization purity.

Observing at longer distances $D$ requires increasing the
observation time $T$ according to $T\propto D^{4/\gamma}$ to obtain
the same number $N$ of sources per solid angle of the galaxy and the
same observation parameter $Q$. Better knowledge of the slope
$\gamma$ of cumulative number counts in the flux range accessible to
the SKA is crucial for observations of distant galaxies.

The existence of simple field modes in a majority of galaxies would
give strong evidence of dynamo action. Failure to detect simple
field structures would suggest that large-scale dynamos are
unimportant in galaxies or that other processes like shearing or
compressing gas flows deform the field lines \citep{beck06}.
Complicated field structures require application of the
reconstruction method, which will become possible for a large number
of relatively nearby galaxies with the SKA, hereby offering
the chance to get good statistics about the magnetic field modes and
their superpositions that may exist in nearby galaxies.

Field recognition may turn out to be also useful at low frequencies.
Equations (\ref{rm_max}) and (\ref{q}) illustrate the advantage of
radio telescopes operating at long wavelengths. For example, LOFAR
will provide a maximum bandwidth of 32~MHz in its lowband
(30--80~MHz) and highband (110--240~MHz). With the large number of
spectral channels available for LOFAR, bandwidth depolarization will
be negligible. The channel width determines the maximum detectable
$\RM$. More severe is the maximum scale in Faraday depth beyond
which an extended source with internal Faraday rotation (i.e.
emission and rotation occur in the same volume) becomes depolarized
\citep{brentjens05}; it varies with $\lambda_{min}^{-2}$, is below
1~rad~m$^{-2}$ in the LOFAR bands and will limit low-frequency $\RM$
measurements of extended sources.

LOFAR is planned to reach an r.m.s. noise (within one hour of
observation time and with 4~MHz bandwidth) of 1--2~mJy/beam area in
the lowband and better than 0.1~mJy/beam area in the highband. This
gives a sensitivity parameter for $\RM$ measurements of
(Eq.~(\ref{a1})) of $A\approx30-50~\mu$Jy~rad~m$^{-2}$ in both
bands, compared to $A\approx3~\mu$Jy~rad~m$^{-2}$ with the SKA at
1~GHz. LOFAR will allow RM to be measured from sources at the
detection limit with errors of $\Delta \RM_{max} \approx
0.005$~rad~m$^{-2}$ in the lowband and $\Delta \RM_{max} \approx
0.1$~rad~m$^{-2}$ in the highband. However, the RM errors will be
much greater under realistic conditions due to the limitations of
correcting for ionospheric Faraday rotation.

The low-frequency array of the SKA should reach an r.m.s. noise of
$1.4~\mu$Jy/beam area within one hour at 200~MHz \citep{carilli04},
hence $A\approx1~\mu$Jy~rad~m$^{-2}$, much better than with the SKA
at higher frequencies and better than with LOFAR, while the maximum
$\RM$ error of $\Delta \RM_{max} \approx 0.1$~rad~m$^{-2}$ is
similar to that using the LOFAR highband.

However, the number density of polarized sources at low frequencies
may be much lower than at high frequencies that would increase the
observation parameter $Q$ (Eq.~(\ref{q})) and hence reduce the
number of $\RM$ points per solid angle of the galaxy. First, the
overall degree of polarization is most probably lower due to Faraday
depolarization effects. Second, the typical degree of polarization
may decrease for more distant sources. Star formation was stronger
in young spiral galaxies, and lobes of radio galaxies were smaller.
Third, internal and external depolarization may significantly reduce
the number of polarized sources (Sect.~\ref{sec:disc}). For planning
$\RM$ surveys at frequencies of 300~MHz and lower, much more needs
to be known about the statistics of polarized sources at low
frequencies.

Another problem at low frequencies is thermal absorption that may
reduce the radio fluxes at frequencies below about 100~MHz for
strongly inclined galaxies with high densities of the ionized gas.

\section{Conclusions}
\label{sec:concl}

%\begin{itemize}

{\em Recognition} of simple structures of regular magnetic fields in
galaxies, characterized by spiral pitch angle, amplitude of regular
field strength, and azimuthal phase, can be reliably performed from
a limited sample of $\RM$ measurements towards polarized background
sources. Applying templates to the field models, we could show that
single and mixed modes can be clearly recognized based on the
$\chi^2$ criterium. The quality of $\RM$ observations can be
characterized by the number $N$ of detected polarized sources within
the solid angle of the galaxy, varying with distance $D$ and
instrumental noise in polarization $\sigma_p$ as $N \propto
\sigma_p^{-\gamma} D^{-2}$, and the ``observation parameter'' $Q$,
which varies as $Q \propto \sigma_p D^{2/\gamma}$ and also depends
on wavelength and bandwidth of the observations. Field recognition
is successful beyond a minimum source number $N^*$ and below a
maximum observation parameter $Q^*$.

The slope $\gamma$ of the cumulative number density counts of
polarized sources with flux densities $P$ ($N \propto P^{-\gamma}$)
is actually known beyond $P=0.5$~mJy from available source counts at
1.4~GHz. Extrapolation to smaller $P$ yields $0.7 \le \gamma \le
1.1$. $\gamma$ still needs to be determined, also at lower and
higher frequencies. For the optimistic case of $\gamma=1.1$,
significant inclination of the galaxy's disk ($i > 25^{\circ}$) and
weak $\RM$ turbulence (due to turbulent magnetic fields and/or
fluctuations of the ionized gas density), a few dozen polarized
sources are already sufficient to reliably recognize a simple field
structure of type ASS, BSS, or QSS. Higher modes (BSS and QSS) are
easier to recognize; i.e., they need less sources or yield smaller
errors of the pitch angle for the same source number compared to the
basic ASS mode. The requirement for a telescope's dynamic range in
polarization (polarization purity) may exceed 30~dB in case of a
flat slope $\gamma$ of the cumulative number density counts of
polarized sources.

Vertical fields do not affect the results of the reconstruction for
galaxies with inclinations $i>15^{\circ}$. The method should not be
applied for galaxies at smaller inclinations where $\RM$ from the
disk field is small and a vertical field may dominate $\RM$.

Strong turbulence or small inclination need more polarized sources
for a statistically reliable recognition. In the case of small
inclination angles, $\RM$ turbulence can get comparable to the $\RM$
amplitude of the regular field and the required number of polarized
sources may become unreasonably large. As the required number of
polarized sources per solid angle also increases for large
inclination angles due to geometry, the recognition method works
best for intermediate inclination angles.

Uncertainties in the position of the galaxy's center of a few
arcseconds are even acceptable for galaxies at 100~Mpc distance if
the field pattern is type ASS. For BSS-type fields accuracies of
better than 2\arcsec\ are needed beyond 50~Mpc distance, and better
than 1\arcsec\ beyond 50~Mpc distance for QSS-type fields.

The application of the recognition method to the available $\RM$
values within the solid angle of the galaxy M~31 is consistent with
the analysis of the diffuse polarized emission if significant $\RM$
contributions intrinsic to the background sources or fluctuating
$\RM$ in the Milky Way foreground are assumed. The regular field of
M~31 is type S0, i.e. axisymmetric spiral in the plane and symmetric
with respect to the plane.

{\em Reconstruction} of the field structure without precognition
needs a large number of sources. A reliable reconstruction of the
field structure is possible for at least 20 $\RM$ values on a cut
along the projected minor axis.

Future RM observations with the SKA at about 1~GHz will extend the
possible targets for field recognition to about 100~Mpc distance,
allow detailed statistical studies of the frequency of field
structures, dependencies on galaxy properties, and test dynamo
against primordial models or other models of field generation. About
60 spiral galaxies can be observed within a distance of 10~Mpc and
$\approx 60000$ spirals within 100~Mpc.

Radio telescopes operating at low frequencies (LOFAR, ASKAP, and the
low-frequency SKA array) may also be useful instruments for field
recognition or reconstruction with the help of $\RM$, if background
sources are still significantly polarized at low frequencies.

%\end{itemize}

\section*{Acknowledgments}

This work is supported by the European Community Framework Programme
6, Square Kilometre Array Design Study (SKADS), contract no 011938,
by DAAD no. A/07/09440 (RS), RFBR no. 06-01-00234, and grant
MK-4338.2007.1 (PF and RS). PF and RS are grateful for financial
support from Prof. Anton Zensus, MPIfR Bonn. We thank Dr. Wolfgang
Reich for careful reading of the manuscript, and thank Prof. Bryan
Gaensler and the anonymous referee for valuable comments. The
simulations were performed on the computer cluster of the IMM
(Ekaterinburg, Russia).
%-----------------------------------------------------

\bibliographystyle{aa} % style aa.bst
\bibliography{biblio}

\end{document}